\documentclass[12pt,preprint] {aastex}

\newcommand\grtsim{\mathrel{\hbox{\rlap{\hbox{\lower2pt\hbox{$\sim$}}}\raise2pt\hbox{$>$}}}}

\newcommand{\whzsr}{W\,Hz$^{-1}$\,sr$^{-1}$}

\shorttitle{Mid-IR spectroscopy of high-$z$ type-2 quasars}
\shortauthors{Mart\'\i nez-Sansigre et al.}

\begin{document}

\title{Mid-infrared spectroscopy of high-redshift obscured quasars}

\author{Alejo Mart\'\i nez-Sansigre$^{1}$, Mark Lacy$^{2}$, Anna Sajina$^{2}$,
Steve Rawlings$^{3}$}
\altaffiltext{1}{Max-Planck-Insitut f\"ur Astronomie, K\"onigstuhl 17,
  D-69117 Heidelberg, Germany; ams@mpia.de} 
\altaffiltext{2}{Spitzer Science Center, California
  Institute of Technology, MS220-6, 1200 E. California Boulevard, Pasadena, CA
  91125, USA; mlacy@ipac.caltech.edu, sajina@ipac.caltech.edu}
\altaffiltext{3}{Astrophysics, Department of Physics, University of
Oxford, Keble Road, Oxford OX1 3RH, UK; sr@astro.ox.ac.uk }

\begin {abstract}
  We present mid-infrared observations of 18 sources from a sample of
  21 $z\sim2$ radio-intermediate obscured quasars. The mid-infrared
  spectra of the sources are continuum dominated, and 12 sources
  show deep silicate absorption with $\tau_{9.7}$$\sim$1-2.  Combining
  mid-infrared and optical spectra, we achieve 86\% spectroscopic
  completeness which allows us to confirm that most (63$^{+14}_{-22}$\%)
   $z\sim2$ radio-intermediate quasars are obscured. The new spectra
  also prove that many high-redshift type-2 quasars do not show any
  rest-frame ultraviolet emission lines. From the 18 individual
  mid-infrared spectra, we classify most of the sources into three
  subsamples: those with hints of the 7.7 and 6.2~$\mu$m polyaromatic
  hydrocarbons (3/18 sources show PAHs, subsample~A), those with an
  excess of emission around 8~$\mu$m but no hint of the 6.2~$\mu$m PAH
  (7/18 cases, subsample~B) and pure-continuum sources with no visible
  excess (4/18 sources, subsample~C). The remaining 4/18 sources have
  spectra that are featureless or too noisy for any features to be
  visible.  In subsample~A, averaging the spectra leads to a
  statistical detection of both 6.2 and 7.7~$\mu$m PAHs over the
  continuum, with the strength of the 7.7~$\mu$m PAH comparable to
  that of submillimetre-selected galaxies (SMGs) at similar
  redshifts. These sources are in a phase of coeval growth of a
  supermassive black hole and a host galaxy.

\end {abstract}


\section{Introduction}

The correlations between galaxy and central supermassive black hole
properties found in the local Universe
\citep{1998AJ....115.2285M,2000ApJ...539L...9F} have led to notions that the
growth of both must be intimately related. A fraction of the gas fuelling star
formation is believed to reach the black hole, which will grow by accretion
during an obscured quasar phase \citep{1999MNRAS.308L..39F}, and is expected
to develop winds or jets that eventually become powerful enough to expel the
gas, preventing any further star formation \citep{1998A&A...331L...1S}. This
feedback process is needed to avoid forming more stars than are seen around
local dormant black holes \citep{2003ApJ...599...38B}, but few compelling
examples of this phase have been reported to date.

Quasars are believed to be viewed as obscured or unobscured depending on the
orientation of the line of sight with respect to the axis of symmetry and the
dusty ``torus'' surrounding the accretion disk \citep{1993ARA&A..31..473A},
yet the population of powerful obscured quasars had remained elusive.  Since
most of the growth of supermassive black holes is obscured by dust, optical
surveys miss this phase completely, and X-ray surveys miss a significant
fraction due to associated gas column. Recently, mid-infrared surveys have
been able to find obscured active galactic nuclei (AGN) in large numbers, but
care must be taken to separate AGN from star-forming galaxies.  This can be
achieved by making use of the particular mid-infrared colours of AGN
\citep[e.g.][]{2004ApJS..154..166L,2006ApJ...640..167A} or using a combination
of mid-infrared and radio data
\citep[e.g.][]{2005Natur.436..666M,2005ApJ...634..169D}.  Indeed, from these
surveys it seems that a large fraction of obscured AGN are undetected in
X-rays \citep[e.g.][]{2005ApJ...634..169D,
  2006ApJ...640..167A,2007AJ....133..186L,2007MNRAS.379L...6M}, and that
mid-infrared selected samples give a more complete picture than X-ray selected
samples.  Some obscured quasars, however, do not show even the narrow emission
lines expected if the central regions are obscured at certain angles by a
torus \citep[see][]{2005Natur.436..666M, 2006MNRAS.370.1479M}, some have
evidence for dust in front of the narrow-line region
\citep{2007ApJ...663..204B}, and radio observations suggest that in some cases
the radio jet is closely aligned with the observer's line of sight
\citep[][and Kl\"ockner et al., in prep.]  {2006MNRAS.373L..80M,
  2007ApJ...667L..17S}, observations which are incompatible with the scenario
of all obscuration being caused by the torus around the accretion disk.

The existence of objects which are not easily explained in the torus
model has led to the suggestion that, in many cases, dust on scales of
kpc is responsible for the obscuration.  Since young stars are born in
dusty regions, it is possible that in these sources the host galaxy is
undergoing an episode of star formation.  Mid-infrared spectroscopy
gives a clean method of looking for star formation in the form of
emission from complex molecules like polycyclic aromatic hydrocarbons
\citep [PAHs, e.g.][]{1985MNRAS.213..777A, 1991MNRAS.248..606R}.
These molecules are excited by ultraviolet photons from young stars,
but destroyed by the hard radiation of an AGN out to $\sim$ 100 pc - 1
kpc scales
\citep[e.g.][]{1991ApJ...379..122V,2004A&A...414..123S}. Thus,
detection of the PAH emission bands can only be attributed to
star-formation and not to excitation by an AGN.

If obscured and unobscured quasars are identical types of objects viewed at a
different orientation, then the properties of their host galaxies, such as the
PAH luminosity, should be identical on average. In a detailed study of
local AGN, \citet {2000A&A...357..839C} found that indeed the PAH luminosities of
Seyfert~1 and Seyfert~2 AGN are statistically indistinguishable, and that the
observed differences between the Seyfert~1 and 2 spectra are due to the
different contrast between the PAHs and the continuum, the latter being weaker
for Seyfert~2s.

In this letter we present the results of mid-infrared spectroscopy of a sample
of radio-intermediate obscured (type-2) quasars,  
which show either narrow lines or no emission lines in their optical
(rest-frame ultraviolet) spectra. 
Throughout
this paper we assume a $\Lambda$CDM cosmology with $h = H_{0} / (100 ~ \rm km
~ s^{-1} ~ Mpc^{-1}) = 0.7$; $\Omega_{\rm m} = 0.3$; $\Omega_{\Lambda} = 0.7$.

\section{Observations and data reduction}

Our sample of radio-intermediate obscured quasars, first presented in
\citet {2005Natur.436..666M}, was observed using the Infrared
Spectrograph (IRS) instrument \citep{2004ApJS..154...18H} on board the
Spitzer Space Telescope \citep[][GO-30634 and archival
data]{2004ApJS..154....1W}.  At $z\grtsim1$, the brightest PAH lines
as well as the silicate features are all redshifted out of the range
covered by the Short-Low module (5.2-14.5~$\mu$m). Therefore, only the
low resolution, long wavelength detector was used (Long-Low), with
both orders, giving coverage in the spectral range
14-38~$\mu$m. However, due to the high noise above 35~$\mu$m, we only
make use of the range 14-35~$\mu$m.  The number of cycles each source
was observed for is shown in Table~\ref{tab:main}. Each cycle involves
two nodded positions, with an individual cycle duration of 120~s at
each position. Note that three sources were not observed due to their
high-redshifts (AMS03, AMS12 and AMS16).

The basic calibration data (BCD) from the Spitzer Science Center IRS pipeline
were then combined by subtracting nodded pairs, and median combining the
pairs. The number of nodded pairs combined in this way varies between 2 and
10, depending on the source (see Table~1). Both mean and median combining were
tried, and for most sources the resulting spectra were almost
indistinguishable.  Mean combining lead to slightly better signal-to-noise
ratio (SNR) spectra in the smooth regions, but median combining generally
proved better at removing spikes of noise.  Given the low SNR for most
of our spectra, median combining was chosen.

The 2-dimensional spectra were cleaned of rogue pixels using
IRSCLEAN\footnote{http://ssc.spitzer.caltech.edu/postbcd/softwarehome.html},
and 1-dimensional spectra were extracted using the optimal extraction in
SPICE$^{1}$. For the optimal extraction, an uncertainty file was first created
from the frame to frame variance of each pixel, using the 440 BCD files of our
own GO program. This uncertainty file was scaled to the correct exposure time
for each source, and is overlayed in Figure~1.  All 18 sources are detected in
the reduced data.

\section{Mid-infrared spectra}

\subsection{Spectral characteristics}\label{sec:spec}

The mid-infrared spectra allow us to obtain spectroscopic redshifts
for a further 8 sources, confirming that 17 out of 21 sources in our
sample lie at $z\geq1.6$. Contamination by pure starbursts is
unlikely, since the mid-infrared spectra are characteristic of AGN-dominated
sources.

Figure~1 shows the 18 individual spectra. Silicate absorption can be
found in most of the sources. In two cases (AMS05, see
Appendix~\ref{sec:opt}, AMS17), the lack of this feature can be
explained due to their high redshift, so the feature is expected to be
centred at and observed wavelength $\lambda_{\rm obs} \geq
37$~$\mu$m. In other sources the low SNR means
it is difficult to see clearly this feature, although hints of it are
seen in the 2D and 1D spectra. In addition, from Figure~1, it can be
seen that AMS01 could be at $z\sim2.1$ and have a shallow silicate
feature, but we do not consider this redshift secure so we do not
quote it in Table~1.

In total, 12/21 sources show no emission lines in the optical (rest-frame
ultraviolet, see also Appendix~A). The mid-infrared spectra of these 12
sources are continuum-dominated and 8 sources show the silicate absorption
feature. These mid-infrared properties are characteristic of AGN.  In fact 6
sources showing no rest-frame ultraviolet lines are at redshifts where the
Ly~$\alpha$ falls in the optical range (that is $z\geq1.7$), and therefore
have no Ly~$\alpha$ detection down to $S_{\rm
  Ly\alpha}\lesssim$$2\times10^{-20}$ W m$^{-2}$. We estimate the absorption
to the broad-line region, at 1216 \AA\, to be $A_{\rm Ly\alpha}\sim60-120$
magnitudes (see Section~\ref{sec:ext}). The extinction towards the narrow-line
region is expected to be significantly lower.  However, the Ly~$\alpha$ line
suffers from resonant scattering, so modest amounts of dust mixed with the
narrow-line region could still severely extinguish the line emission. Indeed,
the Ly~$\alpha$ profile of AMS03 \citep[Figure~6 of][]{2006MNRAS.370.1479M},
is very similar to the simulated spectrum of a young dusty galaxy by
\citet[][Figure~2]{2007ApJ...657L..69L}, which includes resonant scattering.
AMS03 is close to the limit of our sensitivity, so sources where the
narrow-line region is dustier  or where the Ly~$\alpha$ line is
intrinsically less luminous than in AMS03, will not show Ly~$\alpha$ in our
optical spectra.

We find 10 objects with an excess emission around rest-frame 8~$\mu$m, which
at first suggests the 7.7~$\mu$m PAH emission band. These objects are AMS02,
AMS04, AMS06, AMS08, AMS09, AMS11, AMS15, AMS17,AMS18 and AMS21. Care must be
taken, however, since for heavily absorbed sources, there is an apparent
excess at 8~$\mu$m. This is due to the minimum in optical depth between the
absorption features due to silicate, water ice (centred at 6.0 $\mu$m) and
hydrogenated amorphous carbons \citep[at 6.9 and 7.3~$\mu$m, for a detailed
discussion see][]{2002A&A...385.1022S}. At first sight, this excess emission
around 8~$\mu$m can mimic the 7.7~$\mu$m PAH, but it is wider than the
7.7~$\mu$m PAH, and it is not accompanied by the 6.2~$\mu$m PAH.  In 3 sources
(AMS11, AMS17 and AMS21), there is a hint of another emission band
corresponding to the 6.2~$\mu$m PAH.  A mean spectrum of these 3 sources
(hereafter subsample~A) is shown in Figure~2(A), where the 6.2 and 7.7~$\mu$m
PAHs are detected. The mean spectrum of the remaining 7 sources with a maximum
around 8~$\mu$m, but no clear hint of the 6.2~$\mu$m PAH (subsample~B), is
shown in Figure~2(B). Figure~2(C) shows the mean spectrum of subsample~C,
consisting of 4 sources (AMS05, AMS13, AMS14, AMS19) continuum dominated with
clear silicate absorption, but no hint of excess emission around 7.7-8~$\mu$m.
For reference, Figure~2 has the mid-infrared spectra of several low-redshift
ultra luminous infrared galaxies overlayed \citep [ULIRGs, with $L_{\rm
  IR}>10^{12}$ L$_{\odot}$, from][]{2007ApJ...656..148A,2007ApJ...654L..49S}.
These are: IRAS~12514$+$1027 (hereafter IRAS~12514), Mrk~231, Mrk~273 and
Arp~220. The mid-infrared spectrum of IRAS~12514 represents a pure-AGN,
Mrk~231 and Mrk~273 represent AGN-starburst composites, while in the
mid-infrared, Arp~220 is starburst-dominated.

For each subsample, the mid-infrared spectrum of IRAS~12514 was used as a
continuum template. It was normalised using the mean luminosity density of the
6.5-7.5~$\mu$m region, and then subtracted.  The residuals are plotted in
Figure~3, with the spectrum of M~82 \citep{2000A&A...358..481S} overlayed as a
reference for the location of the PAHs. Table~\ref{tab:pah} summarises the
mean PAH luminosities or limits for each subsample.

In subsample~A, there is a detection of the 6.2~$\mu$m PAH (3.4~$\sigma$,
see Table~\ref{tab:pah}) together with the 7.7~$\mu$m PAH (6.4~$\sigma$).
The noise is calculated in the 5.0-8.5~$\mu$m region, (note this is lower than
in the 8.5-11.5~$\mu$m region), and as is mentioned in Table~\ref{tab:pah}, it
is the noise in the 5.0-8.5~$\mu$m region that is used to estimate the
significance of the detections. The uncertainty in the line flux additionally
includes the uncertainty in the continuum subtraction. In subsample~A
the presence of PAHs is an indicator for ongoing starformation in the host
galaxy.

Subsample~B shows an excess around 8~$\mu$m but no hint of the 6.2~$\mu$m
line.  The shape of the excess, and the lack of the 6.2~$\mu$m PAH, strongly
suggest this excess at 8~$\mu$m is continuum and not the 7.7~$\mu$m PAH. 
Thus, in
subsample~B we conclude there is no evidence for PAHs, within our sensitivity,
and that the maximum at 8~$\mu$m is probably continuum. Subsample~C shows no
hint of any PAH or any excess at 8~$\mu$m, and indeed the residuals in
Figure~3 (C) are consistent with noise.  The mid-infrared spectrum of
subsample~C is, within our sensitivity, indistinguishable from that of
IRAS~12514.

\citet {2003MNRAS.338L..19W} obtained an X-ray detection of IRAS~12514, and
estimated the total bolometric luminosity to be $L_{\rm
  bol}\sim$1.6$\times10^{13}$ L$_{\odot}$, with a Compton-thick quasar and a
powerful starburst contributing comparable fractions to $L_{\rm bol}$ (a
source is Compton-thick if the absorbing column density, $N_{\rm H}$, is $\geq
{1/\sigma_{T}}=$$1.5\times10^{28}$ m$^{-2}$, where $\sigma_{T}$ is the Thomson
electron scattering cross-section). The mid-infrared spectrum of IRAS~12514,
however, is AGN-dominated, with a very low value of the 6.2~$\mu$m PAH
equivalent width \citep [see Figure~1 of][]{2007ApJ...654L..49S}. Thus,
although the mid-infrared spectra of samples~B and C show no signs of
starburst activity, vigorous ongoing star-formation in the host galaxy cannot
be ruled out.

\subsection{Derived extinctions}\label{sec:ext}

The spectra are noisy in the silicate region, and from Figure~3 it is
clear that the derived depth of the absorption feature, $\tau_{9.7}$,
will vary vastly depending on the exact wavelength at which it is
measured and the choice of anchor for the continuum \citep [see, e.g.,
][]{2007ApJ...654L..49S}.  However, we can estimate $\tau_{9.7}$ by
comparison with the local ULIRGs. Arp~220 has $\tau_{9.7}=3.3$,
Mrk~231 $\tau_{9.7}=0.8$, and Mrk~273 $\tau_{9.7}=1.8$ \citep[all
three from ][]{2007ApJ...656..148A}, while IRAS~12514 has
$\tau_{9.7}=1.5$ \citep{2007ApJ...654L..49S}. For subsample~A, in the
silicate region, the spectra of Mrk~273 and Mrk~231 represent lower
and upper envelopes, respectively, so $0.8 < \tau_{9.7} < 1.8$. For
subsample~B, IRAS~12514 provides an upper envelope and Arp~220 a lower
one (so $1.5 < \tau_{9.7} < 3.3$), while within the noise, the
spectrum of Mrk~273 is a good fit, so $\tau_{9.7}\sim1.8$. The
spectrum of subsample~C is very close to that of IRAS~12514, so
$\tau_{9.7}\sim1.5$.

We can see that $\tau_{9.7}$ is typically $\sim$1-2. Assuming $A_{\rm
  V}=18.5\times \tau_{9.7}$ \citep{2003ARA&A..41..241D}, we therefore expect
values of $A_{\rm V}$ in the range $\sim$18.5-37.  Using the Milky-Way dust
model of \citet {1992ApJ...395..130P}, one expects the extinction at 1216
\AA\, to be $A_{\rm Ly\alpha}\sim60-120$ magnitudes. This is an estimate of
$A_{\rm Ly\alpha}$ towards the central region (i.e., to the broad Ly~$\alpha$
line).

Following \citet {2006MNRAS.370.1479M}, we assume a gas-to-dust ratio of
$N_{\rm H}=5\times10^{26}$~m$^{-2}$ $\times A_{\rm V}$, which suggests column
densities $\sim9\times10^{27}-2\times10^{28}$~m$^{-2}$.  We compare this to
the absorbing column densities of the AGN in IRAS~12514, Mrk~231 and Mrk~273:
\citet{2005A&A...442..469B} find $N_{\rm H}=7\times10^{27}$ m$^{-2}$ in
Mrk~273, while \citet {2003MNRAS.338L..19W} find $N_{\rm H}>1.5\times10^{28}$
m$^{-2}$ in IRAS~12514.  For Mrk~231, \citet{2004A&A...420...79B} estimate
$N_{\rm H}\sim 2 \times 10^{28}$ m$^{-2}$.  Thus, the low-redshift ULIRGs
whose mid-infrared spectra most resemble our sample are all AGN heavily
absorbed in X-rays, or even Compton-thick.

Indeed, the detection of PAHs in subsample~A suggests the star-forming regions
emitting the PAHs must be shielded from the X-rays, or else the PAHs would be
destroyed \citep{1992MNRAS.258..841V}. The line of sight of the PAHs to the
central source of X-rays is not necessarily the same as our line of sight, but
there is a strong suggestion here that many of the sources in this sample will
be Compton-thick, as was suggested also by \citet{2007MNRAS.379L...6M}.

The mid-infrared spectra can also be used to estimate the mean bolometric
luminosity of each subsample. Assuming the 6.5-7.5~$\mu$m region is dominated
by the AGN continuum, then $L_{\rm bol}\sim10\times$$\lambda
L_{\lambda}$$\sim10^{13}$ L$_{\odot}$ \citep [assuming and unobscured quasar
spectral energy distribution, SED, following][]{1994ApJS...95....1E}.  These
sources are not, of course, unobscured quasars, but the effect of dust at 7
$\mu$m is small: we have estimated the range of $A_{\rm V}$ to be
$\sim$18.5-37. Using the Milky-Way type dust model of
\citet{1992ApJ...395..130P}, the expected transmission at 7~$\mu$m is
$\sim$60-75\% for this range of $A_{\rm V}$, while it is $\sim$50-55\% for
Small-Magellanic Cloud type dust.  Hence, the bolometric luminosity is
underestimated by at most 50\%, which is smaller than or comparable to the
intrinsic uncertainty in using the Elvis et al. (1994) conversion \citep
[which is a factor of $\sim2$, see also ][]{2006ApJS..166..470R}.  Note that
values of $L_{\rm bol}$$\grtsim10^{13}$ L$_{\odot}$ are also estimated for the
sample of \citet{2007MNRAS.379L...6M}, which was selected using very similar
criteria as well as broad-band SED fitting.

\subsection{Possible selection bias?}

A source of concern is whether there is a selection effect in our sample: by
being selected at 24~$\mu$m, it could be biased in favour of $z\sim2$ sources
with strong PAHs. To a certain extent this is inevitable, but we show in this
section how the effect is minimal. In addition, a look at the spectra in
Figure~1 shows such a selection effect has not heavily affected us.

At $z=2$, the 7.7~$\mu$m PAH is almost perfectly centred on the MIPS 24~$\mu$m
band. We estimate the
mean contribution to the 24~$\mu$m flux density of a $L_{7.7}=5\times10^{10}$
L$_{\odot}$ PAH (from Table~\ref{tab:pah}) at $z=2$ to be $\sim$100~$\mu$Jy. This can clearly be
important for a sample whose flux density limit is 300~$\mu$Jy. Thus, at the lower
flux density end, our sample could be biased in favour of sources with PAHs, yet we
expect at most  $\sim30$\% contribution of the PAH to the 24~$\mu$m flux
density, and for sources with $S_{24}>650$~$\mu$Jy, the contribution of the
PAH will be $\lesssim15$\% and therefore smaller than or comparable to the
uncertainty in the flux density.

Amongst the faint sources, we do see cases where the sources have been selected
due to the maximum around 8~$\mu$m falling in the 24~$\mu$m band (e.g. AMS02,
AMS15, see Figure~1). Yet, as we have discussed in Section~\ref{sec:spec},
without detection of the 6.2~$\mu$m line, we believe this 8~$\mu$m excess to
be continuum and not the 7.7~$\mu$m PAH.  We have only identified PAHs in
AMS11, AMS17 and AMS21. 

In the case of AMS11, the peak of the 7.7~$\mu$m PAH is at 20~$\mu$m, but the
MIPS-24~$\mu$m band has a spectral response curve with a full-width half
maximum between 20.8 and 26.1~$\mu$m \citep{2004ApJS..154...25R}. Hence, at
most half of the 7.7~$\mu$m PAH flux falls in the band, so the PAH
contribution is expected to be $\sim50$ $\mu$Jy. AMS11 has a 24~$\mu$m flux density,
$S_{24}=442$ $\mu$Jy \citep[24~$\mu$m flux densities from ][]{2005Natur.436..666M}, so
subtracting the PAH contribution one expects $S_{24}\sim 392$ $\mu$Jy, enough
to be inside our sample anyway.  AMS17 is at $z=3.137$, so only the 6.2~$\mu$m
PAH falls in the 24~$\mu$m band.  It has $S_{24}=1134$ $\mu$Jy: clearly the
PAHs have no effect in the inclusion of this source in our sample. AMS21 is at
$z=1.8$ and the PAH falls entirely within the 24~$\mu$m band, but with
$S_{24}=720$ $\mu$Jy the PAH-subtracted flux density is $S_{24}\sim620$ $\mu$Jy,
enough to be included. A look at AMS21 in Figure~1 suggests the continuum to
be slightly lower than this estimate, around $S_{24}\sim 400$ $\mu$Jy, but
still bright enough to be included in the sample.

Overall, the spectra shown in Figures~1 and 2 show our sample is not
contaminated by pure starbursts, and in any case the radio properties suggest
all our sources contain AGN: the sources have radio luminosities,
$L_{1.4}\sim10^{24}$ W Hz$^{-1}$ sr$^{-1}$, several have flat- or
gigahertz-peaked radio spectra \citep{2006MNRAS.373L..80M}, and many have
radio cores detected (Kl\"ockner et al., in prep.).

\section{The quasar fraction revisited}\label{sec:fraction}

We use our new mid-infrared spectra (and the new optical spectra in
Appendix~\ref{sec:opt}) to revisit the quasar fraction at $z\sim2$,
following \citet{2005Natur.436..666M}. Above $z=1.7$, Ly~$\alpha$ is
potentially visible in optical spectroscopy, while in our mid-infrared
spectra, the 9.7~$\mu$m silicate absorption feature is visible in the
approximate range $1.4 \lesssim z \lesssim 2.6$. Thus, our
spectroscopic completeness to both ``host-obscured'' and
``torus-obscured'' quasars should now be close to 100\% in this range.
Above $z\sim2.6$ we are no longer complete to sources showing no
emission lines in the optical spectra. Thus, we estimate the quasar
fraction at $1.7\leq z\leq2.6$, where we expect to have close to 100\%
spectroscopic completeness. We note that objects with no narrow lines
in the optical spectrum and with shallow silicate absorption features
could still be missing from our census (e.g., AMS01 might be at
$z\sim2.1$ as discussed in Section~\ref{sec:spec}).

We begin by estimating the number of type-1 quasars following our 24~$\mu$m
and radio criteria.  This is estimated by using the
\citet{2003A&A...408..499W} luminosity function (assuming pure luminosity
evolution, PLE), together with a typical type-1 SED \citep [essentially flat
in $\nu L_{\nu}$ in the infrared, from][]{1995MNRAS.272..737R} are used to
estimate the number of $1.7 \leq z \leq 2.6$ type-1 quasars meeting our
24~$\mu$m cut. The optical-to-radio correlation of \citet{2003MNRAS.346..447C}
is then used to see what fraction of these type-1s would also meet the radio
cuts.  We estimate the errors by changing from PLE, to pure density evolution
(PDE). We also vary the mid-infrared spectral index (where $L_{\nu} \propto
\nu^{-\alpha_{\rm MIR}}$) of the typical type-1 SED, assumed to be
$\alpha_{\rm MIR}=1$, in the range $0.87 \leq \alpha_{\rm MIR} \leq 1.13$, and
the radio spectral index, assumed to be $\alpha_{\rm rad}=0.8$, in the range
$0.5 \leq \alpha_{\rm rad} \leq 1$. The errors are then added in quadrature,
with the difference between PDE and PLE considered as a $\pm1\sigma$ error,
while the range spanned by the spectral indices is considered to be
$\pm2\sigma$. In an area of 3.8 deg$^{2}$, we predict 6.0$^{+2.3}_{-1.4}$ such
type-1 quasars.

Our sample includes 4 objects with spectroscopic redshifts $1.7 \leq z \leq
2.6$ from rest-frame ultraviolet lines in the optical spectra. A further 6
objects have $1.7 \leq z \leq 2.6$ from the mid-infrared spectra (see
Table~1). Figure~4 shows the normalised posterior probability distribution of
the quasar fraction, $q$, given our data, given the modelled number of type-1
quasars, and assuming a prior probability distribution for the quasar fraction
flat in the entire possible parameter range of $0 \leq q \leq 1$ (again,
following Mart\'\i nez-Sansigre~et al.  2005). The solid blue line shows the
quasar fraction when only the type-2 quasars with optical redshifts are used
(4 type-2s for a predicted 6.0 type-1s): it has a modal value of 0.60 and 68\%
of the area lies in the region with $q \geq 0.51$ (note that, for this curve,
the errors are so large that the probability of $q>1.0$ is greater than zero).
The dashed red line shows the posterior distribution for $q$ when all 10
type-2 quasars are considered: it has a modal value of 0.37 and the region $
0.23 \leq q \leq 0.59$ encompasses 68\% of the area.

We summarise the derived quasar fraction as $q=0.60^{+0.40}_{-0.09}$ for
type-2 quasars showing narrow-lines only, and $q=0.37^{+0.22}_{-0.14}$ for all
type-2 quasars with spectroscopic redshifts. The 'receding-torus' model of
\citet [][marked 'S05']{2005MNRAS.360..565S}, was derived from type-1 and
type-2 AGN showing emission lines and should only be compared to the solid
blue line of Figure~4. His model predicts a value of $q=0.63$ at high AGN
luminosities, in good agreement with our results.

We stress that the uncertainties are still large. For example, if instead we
derive the quasar fraction at all $z\geq1.7$, we estimate 6.7$^{+3.0}_{-2.2}$
type-1 quasars, and the two quasar fractions as $q=0.45^{+0.24}_{-0.16}$
(using only type-2 quasars with narrow lines and $z_{\rm spec}\geq 1.7$) and
$q=0.32^{+0.19}_{-0.12}$ (using all type-2 quasars with $z_{\rm spec}\geq
1.7$). However, for the range $1.7\leq z\leq 2.6$, where our spectroscopic
completeness is highest, our best estimate of the radio-quiet quasar fraction
is $q=0.37^{+0.22}_{-0.14}$ or, equivalently, $63^{+14}_{-22}$\% of
radio-quiet quasars are obscured.

\section{Comparison with other high-redshift samples}

Searches for PAHs in high-redshift (unobscured) quasars have proven difficult:
\citet{2007A&A...468..979M} observe a sample of 25 optically bright $z\sim2-3$
quasars, find no detection even in their stacked spectrum, and derive an upper
limit on the luminosity of the 7.7~$\mu$m PAH, $L_{7.7}\leq 4.6\times10^{10}$
L$_{\odot}$. The sources comprising the sample of Maiolino et~al. (2007) have
bolometric luminosities significantly larger than those of our own sample,
making the PAHs difficult to see over the strong continuum. However, this is a
good comparison sample as it is unbiased in terms of the star-formation rates
in the host galaxies.

\citet{2007ApJ...661L..25L} do find a detection in the ``Cloverleaf'' quasar,
with $L_{7.7}=7.6\times10^{10}$ L$_{\odot}$, although this is a biased case
since it was observed because of it being hyper-luminous in the far-infrared
(even after correction for gravitational magnification).

In Figure~5, the PAH strengths and estimated
bolometric luminosities of subsamples~A, B and C are compared to those
of the unobscured quasars mentioned above and the sample of
submillimetre-selected galaxies (SMGs) of
\citet{2007ApJ...660.1060V}. The SMGs represent the most powerful
starbursts known, and have their bolometric luminosities approximated
to their total infrared luminosities
\citep[from][]{2007ApJ...660.1060V}.  Detection of the PAH is not an
artefact of any selection criteria other than SMGs being galaxies with
vigorous star-formation.

Our sample of obscured quasars is around the break in the LF
\citep{2004MNRAS.349.1397C}, so these sources represent the bulk of
the energy density (or accreted mass density) of SMBHs. However, they
are radio-intermediate (with $L_{1.4 \rm GHz} \sim 10^{24}$ \whzsr),
and therefore are more rare in space density than genuinely radio-quiet
obscured quasars. In addition, this radio selection could possibly
affect the observed properties, something that must be kept in
mind. If the quasars were obscured solely by the torus of the unified
schemes, the luminosity of the PAHs from star-formation in the quasar
host galaxies should be similar for unobscured and obscured
quasars.

In our sample, emission from PAHs is detected in 3 out of 18 objects
observed (subsample~A).  For this subsample, the inferred values of
$L_{7.7}$ are similar to those of the SMGs, and only slightly lower
than the value of the Cloverleaf.   
The limits inferred for non-detections (both for subsample~C and the
unobscured quasars of Maiolino et~al., 2007) are similar to the values of the
detections, so little can be said about any differences in PAH properties
between samples.

The obscured quasars with detections of PAHs, have similar values of $L_{7.7}$
to the SMGs, which are powerful starbursts.  The actual conversion from
$L_{7.7}$ to a star-formation rate (SFR) is more uncertain, since for higher
luminosity starbursts $L_{7.7}$ does not increase linearly with the
far-infrared luminosity \citep{2001A&A...367L...9H, 2001A&A...379..823K,
  2002ApJ...576..159D}.  Although it is difficult to estimate the
star-formation rates (SFRs) from $L_{7.7}$, the values of $L_{7.7}$ estimated
for obscured quasars are similar to those of SMGs (see
Figure~5), which have typical SFRs $\grtsim$1000 M$_{\odot}$
yr$^{-1}$ (of all stars, integrated over a Salpeter initial mass function)
estimated from their rest-frame FIR luminosities.  These values of $L_{7.7}$
are, in addition, larger than the limits inferred for lower-redshift X-ray
absorbed quasars \citep{2006ApJ...642...81S}, or the values of $L_{7.7}$ found
in nearby ``PG'' quasars \citep{2006ApJ...649...79S}, which have lower
far-infrared luminosities and thus inferred SFRs ($\lesssim100$ M$_{\odot}$
yr$^{-1}$) than the SMGs.

\section{Conclusions}

Two main conclusions can be extracted from this work.  Firstly, we
have found that the mid-infrared spectra are continuum dominated, and
12 out of 18 spectra show a deep silicate absorption feature (with
$\tau_{9.7}$$\sim$1-2).  We have confirmed spectroscopically the
existence of sources with mid-infrared spectra characterstic of
heavily obscured quasars, but that do not show rest-frame ultraviolet
emission lines (observed in the optical). The number of these sources
is similar to the number of obscured quasars that do show ultraviolet
narrow emission lines. Using the new mid-infrared redshifts, we
confirm that the population of radio-intermediate obscured (type-2)
quasars outnumbers the unobscured population (type-1), and that
63$^{+14}_{-22}$\% of $z\sim2$ radio-intermediate quasars are obscured.
This had already been suggested in \citet{2005Natur.436..666M}, and in
this paper we confirm the result spectroscopically: amongst
radio-intermediate sources, the population of obscured quasars is
responsible for the bulk of the accretion onto supermassive black
holes.

Secondly, we have found that obscured quasars are sometimes hosted by
galaxies undergoing vigorous star formation, with detectable PAHs comparable
in strength to those found in SMGs. The accompanying SFR is expected to be
very large ($\grtsim$100 M$_{\odot}$ yr$^{-1}$), so that it can form a
significant fraction of an $L^{*}_{\rm K}$ galaxy in $\sim10^{7-8}$ yr.  The
dust and gas fuelling this star formation are likely to be responsible for at
least part of the obscuration in some of the sources, with the accreting black
hole deeply embedded.  These properties are those expected from an obscured
phase of black hole growth \citep{1999MNRAS.308L..39F}, and could explain why
in some obscured quasars the rest-frame ultraviolet lines are not detectable.

\section{ Acknowledgments.}  

We thank Dave Bonfield for help with the WHT observations, Henrik Spoon for
access to the IRS guaranteed time observer spectra and for useful comments,
Kate Brand, Hans-Rainer Kl\"ockner and
Dan Smith for communicating results prior to publication, and Roberto Maiolino
for discussions. We also thank the anonymous referee for useful comments. This work was
partially supported by grants associated with Spitzer programs GO-20705 and
GO-30634, and is based on observations made with the Spitzer Space Telescope,
which is operated by the Jet Propulsion Laboratory, California Institute of
Technology under a contract with NASA. The WHT is operated on the island of La
Palma by the Isaac Newton Group in the Spanish Observatorio del Roque de los
Muchachos of the Instituto de Astrof\'\i sica de Canarias.

%\bibliographystyle{apj.bst} 
%\bibliography{/home/martinez/bibliography/aamnem99,/home/martinez/bibliography/references_database,/home/martinez/bibliography/irsref} 

\clearpage

\appendix

\section {New optical spectra}\label{sec:opt}

For AMS18, the redshifts derived from optical and mid-infrared spectroscopy
disagree completely \citep [compare Table~\ref{tab:main} to Table~1
of][]{2006MNRAS.373L..80M}, and AMS15 had not been observed in optical
spectroscopy. AMS03 was found to have an interesting source nearby: in imaging
data, the source
was point-like in K, but galaxy-like in R, suggesting a reddened quasar.  For
these reasons the 3 objects were observed using the ISIS spectrograph at the
William Herschel Telescope in July 2007.

In the spectrum of AMS18 reported in \citet {2006MNRAS.373L..80M}, three very
weak lines were visible, at 4680, 5652 and 7527 \AA, yet our new spectrum
shows no hint of these three lines. Thus, it is most likely that they were not
real and the redshift reported in \citet {2006MNRAS.373L..80M} is incorrect.
Thus, we favour the redshift of $z=1.6$ from mid-infrared spectroscopy, and
this value is quoted in Table~1.

In AMS15 we find no evidence for any optical lines, in either the blue or red
spectra.  For AMS03, the slit was placed at a different position angle (31
degrees) to go through the reddened quasar. In the 1D spectrum of AMS03, we
see the same double-peaked emission as reported in \citet
[][Figure~6]{2006MNRAS.373L..80M}. The optical spectrum of the candidate
reddened quasar looks like that of an elliptical galaxy at $z=0.7$, and no
obvious emission lines are visible.

In addition, for AMS05, a new optical spectrum and imaging (using both narrow
and broad bands) suggest the Ly~$\alpha$ line had been missidentified as C~IV,
and the object is actually at $z=2.850$ rather than $z=2.017$ (D. Smith,
private communication).

Comparing optical and mid-infrared spectral properties, and radio properties,
we see that none of the sources with flat radio spectra show narrow lines in
their optical spectra (AMS07, AMS15, AMS18 and AMS19, see
Table~\ref{tab:main}). This is important as, from unified schemes, one expects
flat-spectrum radio sources to have broad lines (except in the case of
blazars, where the beamed continuum can outshine the broad-line region). Thus,
if these sources are obscured, the obscuration must occur due to dust on a
larger scale (not due to the torus) and so the narrow lines, as well as the
broad lines, are expected to be obscured. Alternatively, the dust might be on
a small scale similar to that of the torus, but with a covering fraction
$\sim1$, so that the narrow-line region cannot be excited by photons from the
central engine.

\clearpage

\begin{table}
\caption{ Measured and derived properties of our sample of obscured quasars.}
\begin{center}
\begin{tabular}{lllrccccc}
\hline
\hline
Name & RA & Dec & $z_{\rm}^{a}$  & Optical$^{b}$ & Radio$^{c}$ &  $N_{\rm LL1}^{d}$ & $N_{\rm LL2}^{d}$ & Sub$-$  \\

& (J2000)  & & & Spectrum& Spectrum & &  &sample$^{e}$ \\
\hline
AMS01 & 17 13 11.17 & +59 55 51.5  & - & B & SS & 10 & 10 &  N.F. \\
AMS02 & 17 13 15.88 & +60 02 34.2  & 1.8 & B & SS & 10 & 10 & B  \\
AMS03 & 17 13 40.19 & +59 27 45.8  & 2.698 &NL & SS & 0 & 0 & N.O. \\
AMS04 & 17 13 40.62 & +59 49 17.1  & 1.782 &NL & SS & 10 & 10 &B \\
AMS05 & 17 13 42.77 & +59 39 20.2  & 2.850$^{f}$ &NL & GP & 3 & 3 & C \\
AMS06 & 17 13 43.91 & +59 57 14.6  & 1.8 & B & SS & 6 & 6 & B  \\
AMS07 & 17 14 02.25 & +59 48 28.8  & -& B  & FS &  10 & 10 & N.F. \\
AMS08 & 17 14 29.67 & +59 32 33.5  & 1.979 & NL& - & 9 & 9 &  B \\
AMS09 & 17 14 34.87 & +58 56 46.4  & 2.1 & B & SS &8 & 6 & B \\
AMS10 & 17 16 20.08 & +59 40 26.5  & - & B & SS &10 & 10 & N.F. \\
AMS11 & 17 18 21.33 & +59 40 27.1  & 1.6 & B & SS &10 & 10 & A \\
AMS12 & 17 18 22.65 & +59 01 54.3  & 2.767 & NL & SS & 0 & 0 & N.O.   \\
AMS13 & 17 18 44.40 & +59 20 00.8  & 1.974 & NL & SS & 3 & 2 & C \\
AMS14 & 17 18 45.47 & +58 51 22.5  & 1.794 & NL & SS & 3 & 2 &C \\
AMS15 & 17 18 56.93 & +59 03 25.0   & 2.1 & B & FS & 10 & 10 & B \\
AMS16 & 17 19 42.07 & +58 47 08.9   & 4.169 & NL & GP &  0 & 0 & N.O.  \\
AMS17 & 17 20 45.17 & +58 52 21.3  & 3.137 & NL & SS& 8 & 6 & A \\
AMS18 & 17 20 46.32 & +60 02 29.6   & 1.6$^{f}$ & B& FS & 10 & 10 & B \\
AMS19 & 17 20 48.00 & +59 43 20.7 & 2.3 & B & FS & 5 & 5&   C \\
AMS20 & 17 20 59.10 & +59 17 50.5 & - & B & GP & 10 & 10 &  N.F. \\
AMS21 & 17 21 20.09 & +59 03 48.6  & 1.8 & B& SS& 10 & 10  &A \\
\hline
\hline
\end{tabular}
\end{center}
\tablenotetext{a}{The redshifts
  are all spectroscopic. Redshifts with 3 decimal places are from
  optical spectroscopy, from \citet{2006MNRAS.370.1479M} except AMS05 and
  AMS18 as discussed in Appendix~\ref{sec:opt}, while those with only 1 decimal place are from mid-infrared
  spectroscopy. }
\tablenotetext{b}{ Summary of optical spectroscopy properties. B stands for blank spectrum, NL for narrow lines. }
\tablenotetext{c}{ Summary of radio spectral properties from \citet {2006MNRAS.373L..80M}. SS stands for steep spectrum, FS for flat spectrum and GP for gigahertz-peaked source.}
\tablenotetext{d}{N$_{\rm LL1}$ and N$_{\rm LL2}$ are the
  number of cycles used in each of the LL1 and LL2 orders respectively. All
  cycles have a duration of 120~s.  AMS03, AMS12 and AMS16
  where not observed  with IRS due to their high redshift.}
\tablenotetext{e}{ For AMS01, AMS07, AMS10 and AMS20, it was not possible to identify
  securely any features and a redshift, so they are labelled as having no
  features (N.F.). Obviously, AMS03, AMS12 and AMS16
  cannot be classified into any subsample, since they were not observed
  (N.O.). For the rest of the sources, the letter of which subsample each source belongs is quoted.}
\tablenotetext{f}{As discussed in Appendix~\ref{sec:opt}, these two redshifts
  have been revised. The new redshift for AMS05 is from optical spectroscopy
  (D. Smith, private communication), while for AMS18 it is from mid-infrared
  spectroscopy.   }
\label{tab:main}
\end{table}

\clearpage

\begin{table}
  \caption{ Estimated PAH strength and significance}
  %{\footnotesize}
  \begin{center}
    \begin{tabular}{cccccc}
      \hline
      \hline
      Subsample & $L_{\rm bol}$ & $L_{6.2}$$^{a}$ & Signif.$^{b}$& $L_{7.7}$$^{a}$ &  Signif.$^{b}$ \\
      & /$10^{13}$L$_{\odot}$ & /$10^{10}$L$_{\odot}$ & /$\sigma$&  /$10^{10}$L$_{\odot}$ & /$\sigma$  \\
      \hline
      A & 0.9& 2.2$\pm$1.8& 3.4 & 4.6$\pm$1.8&6.4 \\
      %A2 & 0.9& 2.4$\pm$1.8& 3.6 & 3.5$\pm$1.8&5.0 \\
      B & 0.8& $\leq$5.4& $\leq$3 & $\leq$5.4& $\leq$3 \\
      C & 2.8& $\leq$6.5& $\leq$3 & $\leq$6.5 &$\leq$3 \\
      \hline
      \hline
    \end{tabular}
  \end{center}
  \tablenotetext{a}{ The estimated uncertainties include both the uncertainty in removing the continuum and the noise of the residual. }
  \tablenotetext{b}{ For detections we quote the significance of the lines
    above the noise of the residual, the uncertainty in removing the continuum
    is not included. For non-detections, we quote a 3~$\sigma$ limit.}
  \label{tab:pah}
\end{table}

\clearpage

\begin{figure*}
  \includegraphics{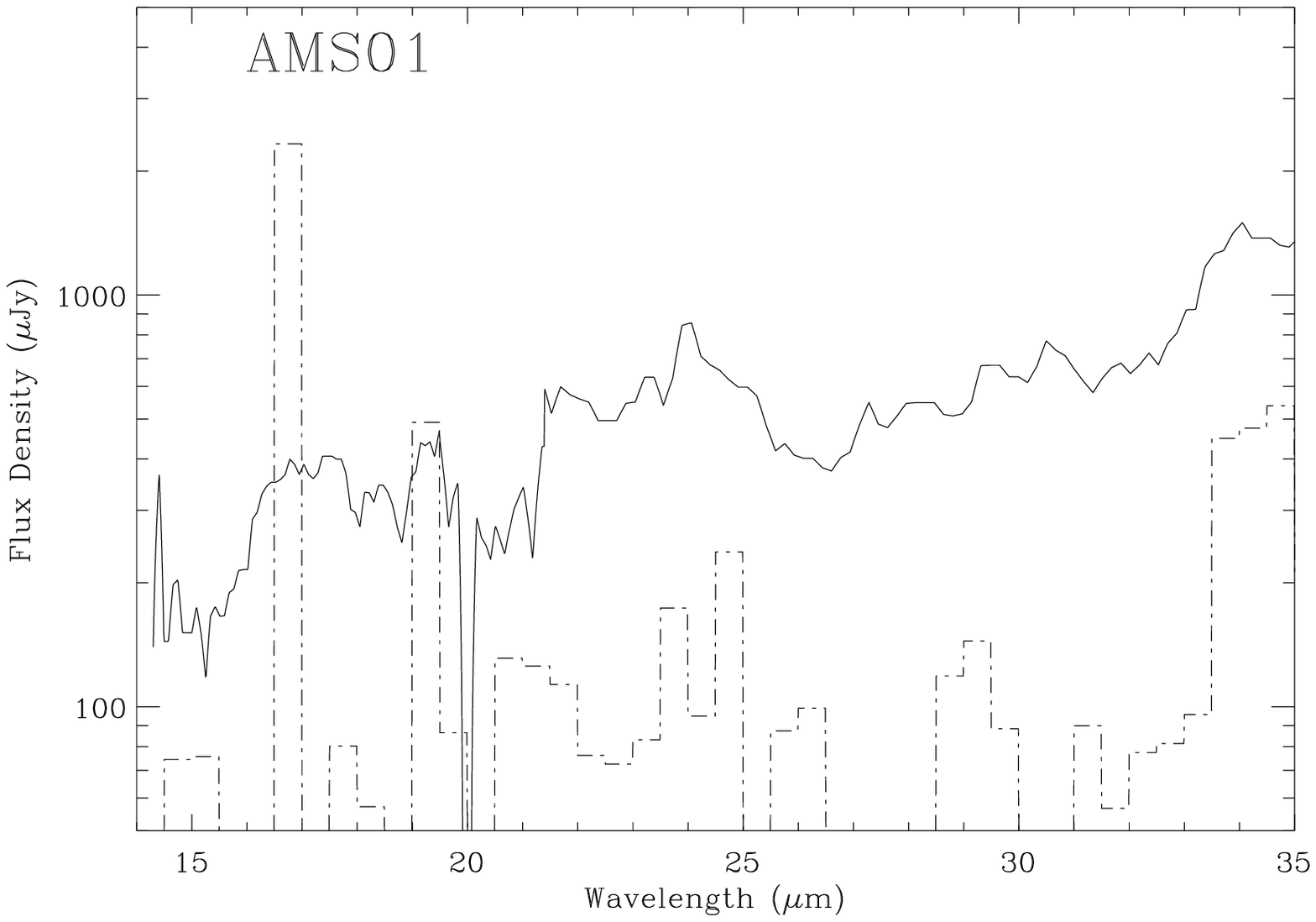} 
  
  \includegraphics{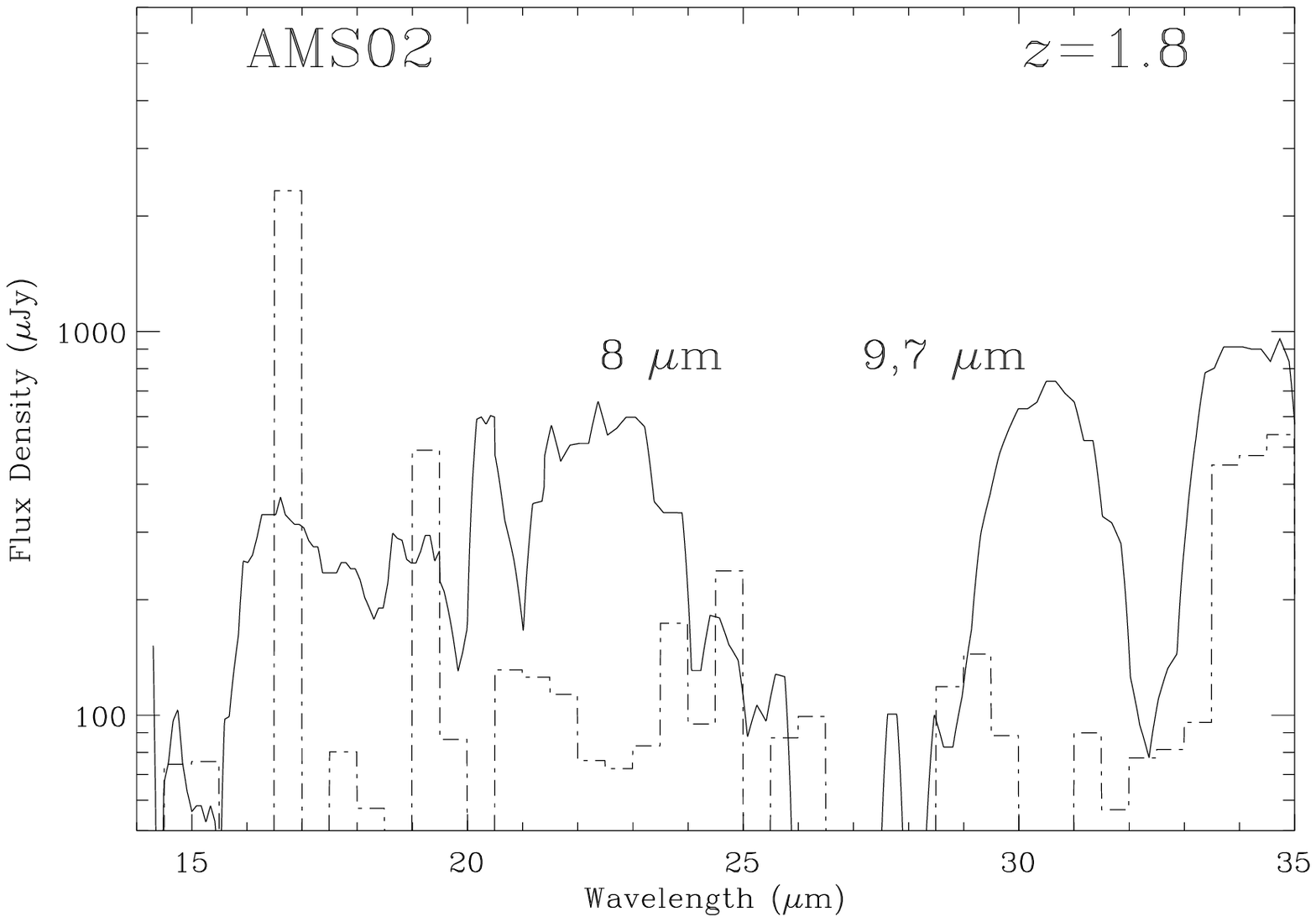} 
  
\includegraphics{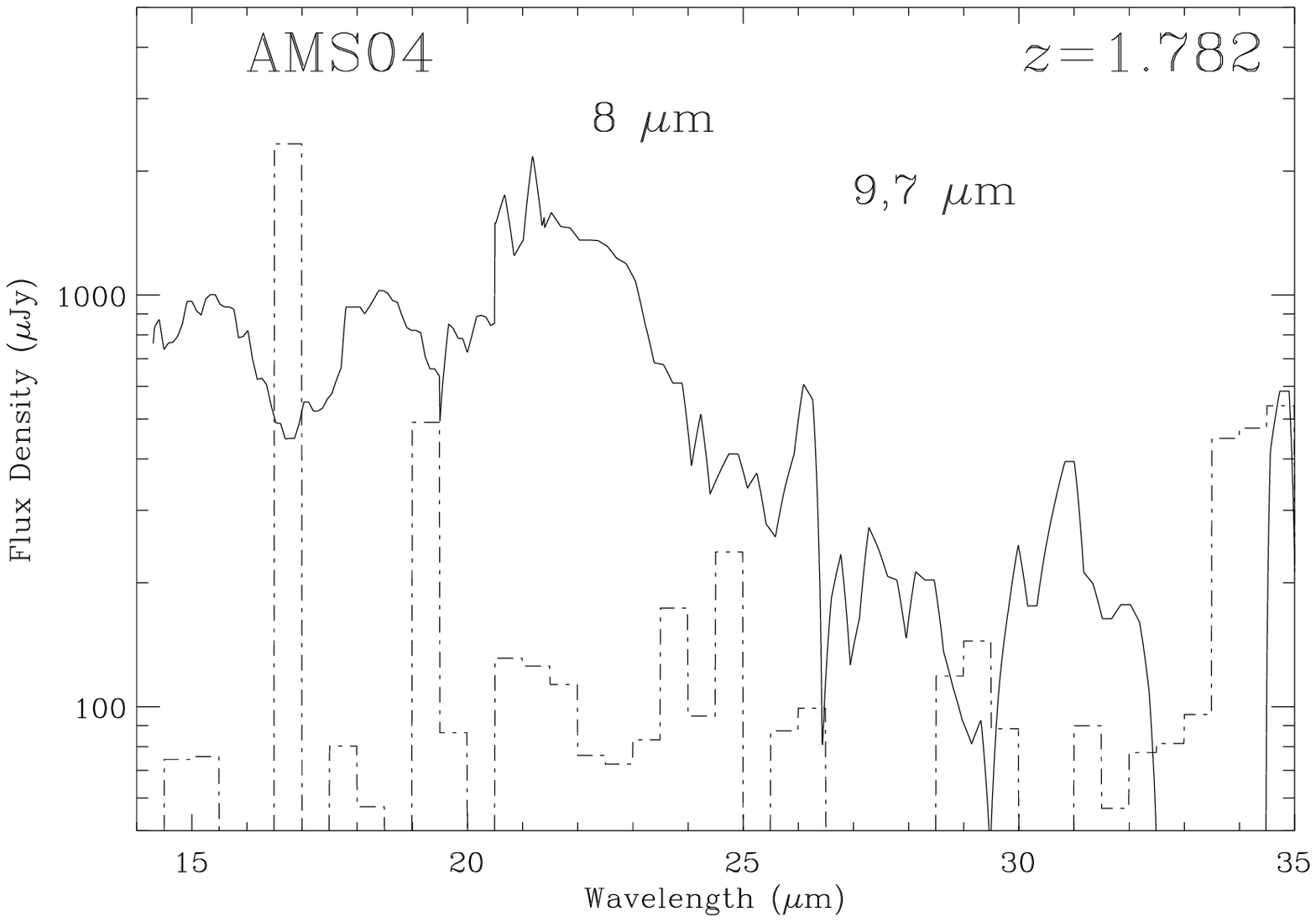} 

\includegraphics{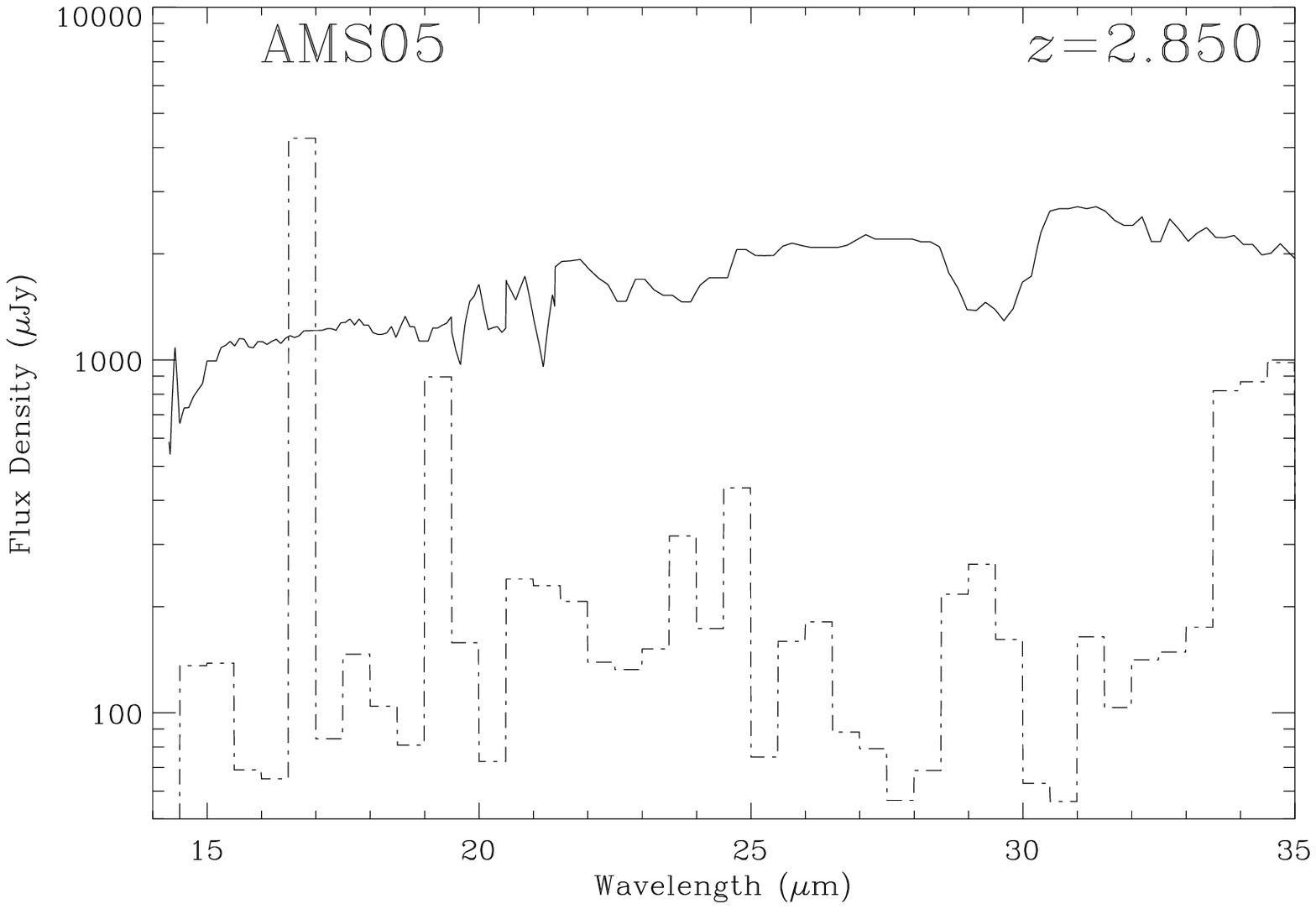} 

\includegraphics{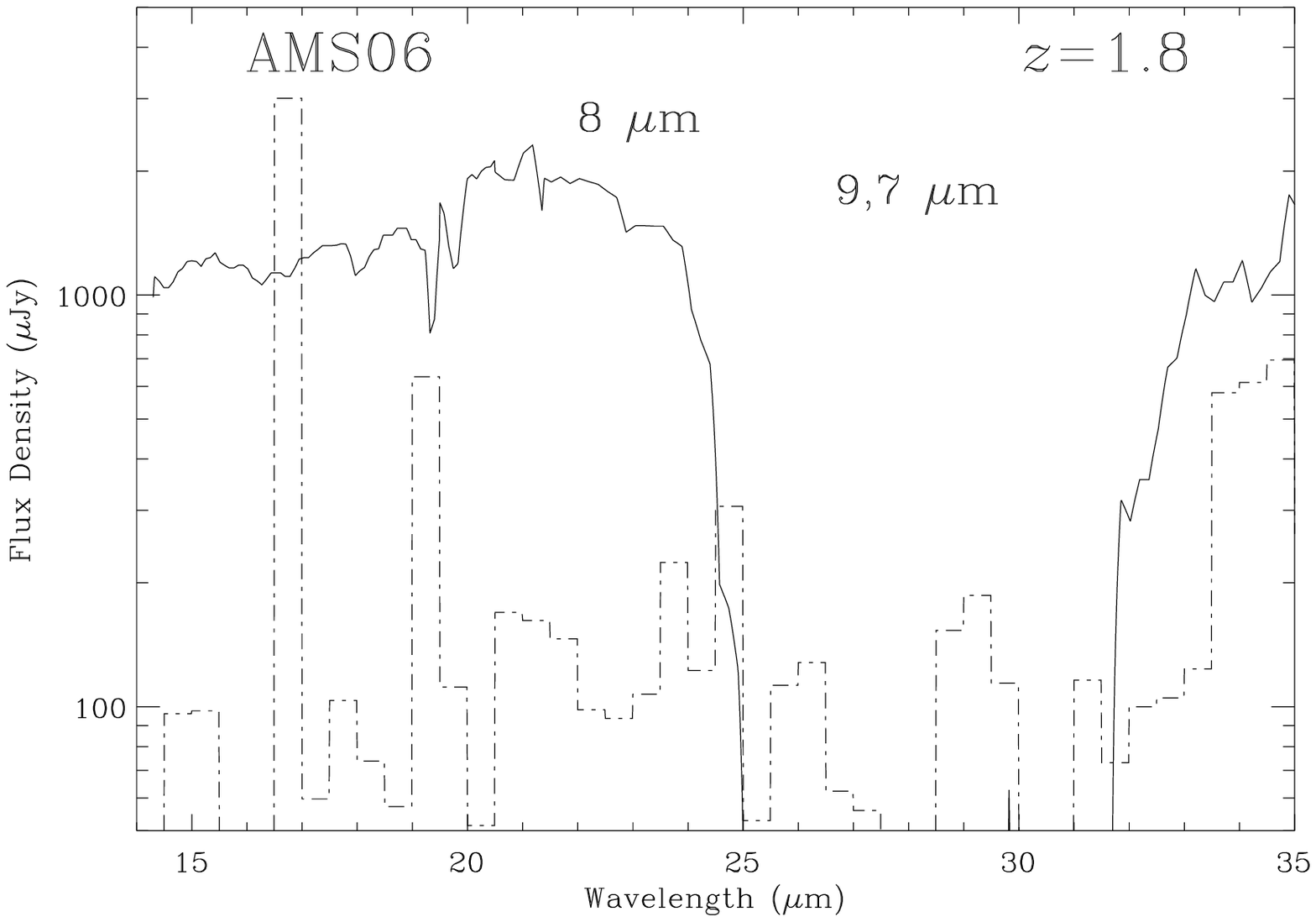} 

\includegraphics{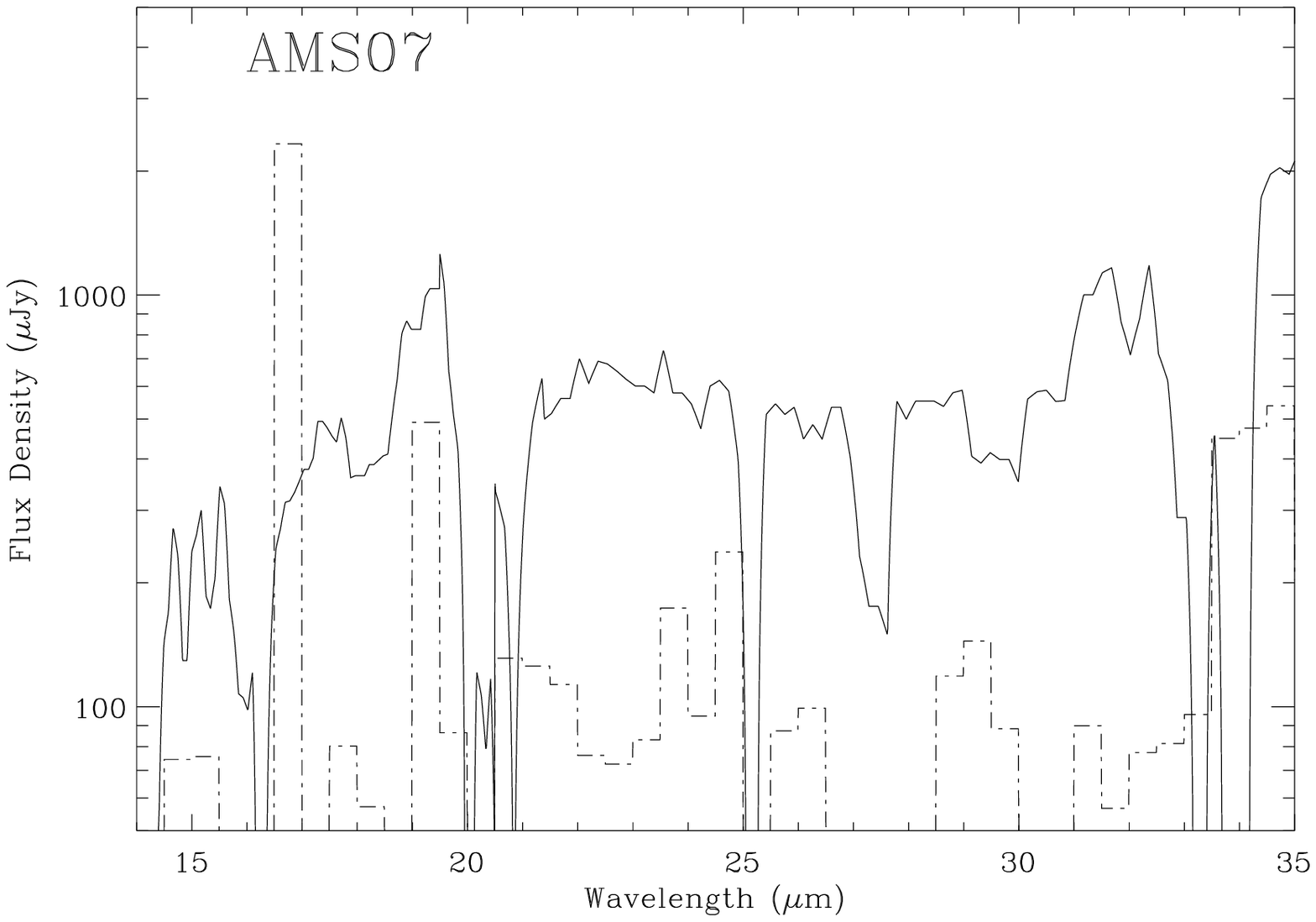} 
{ \caption{ }}
\end{figure*}

\clearpage

\addtocounter{figure}{-1}
\begin{figure*}

\includegraphics{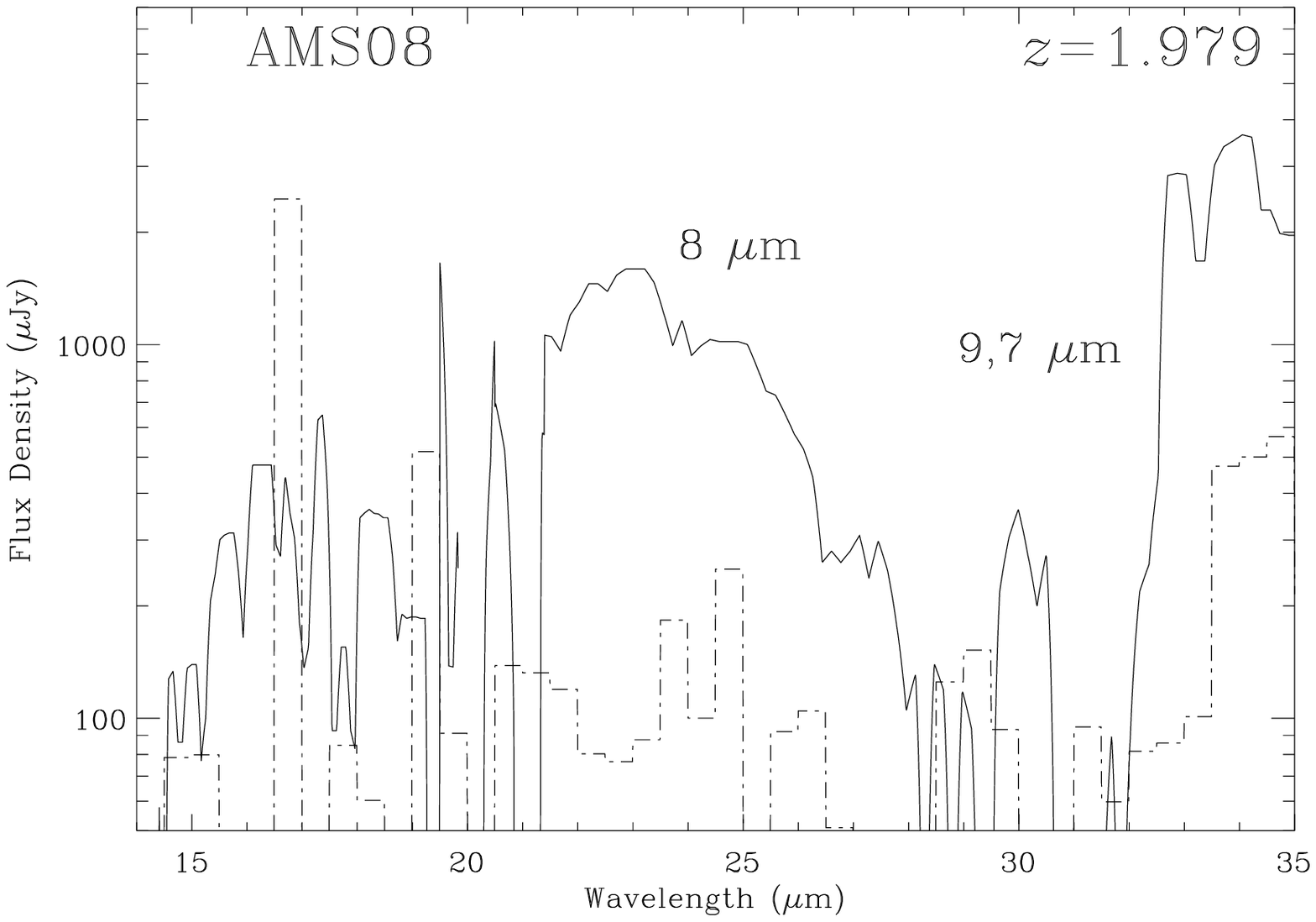} 

\includegraphics{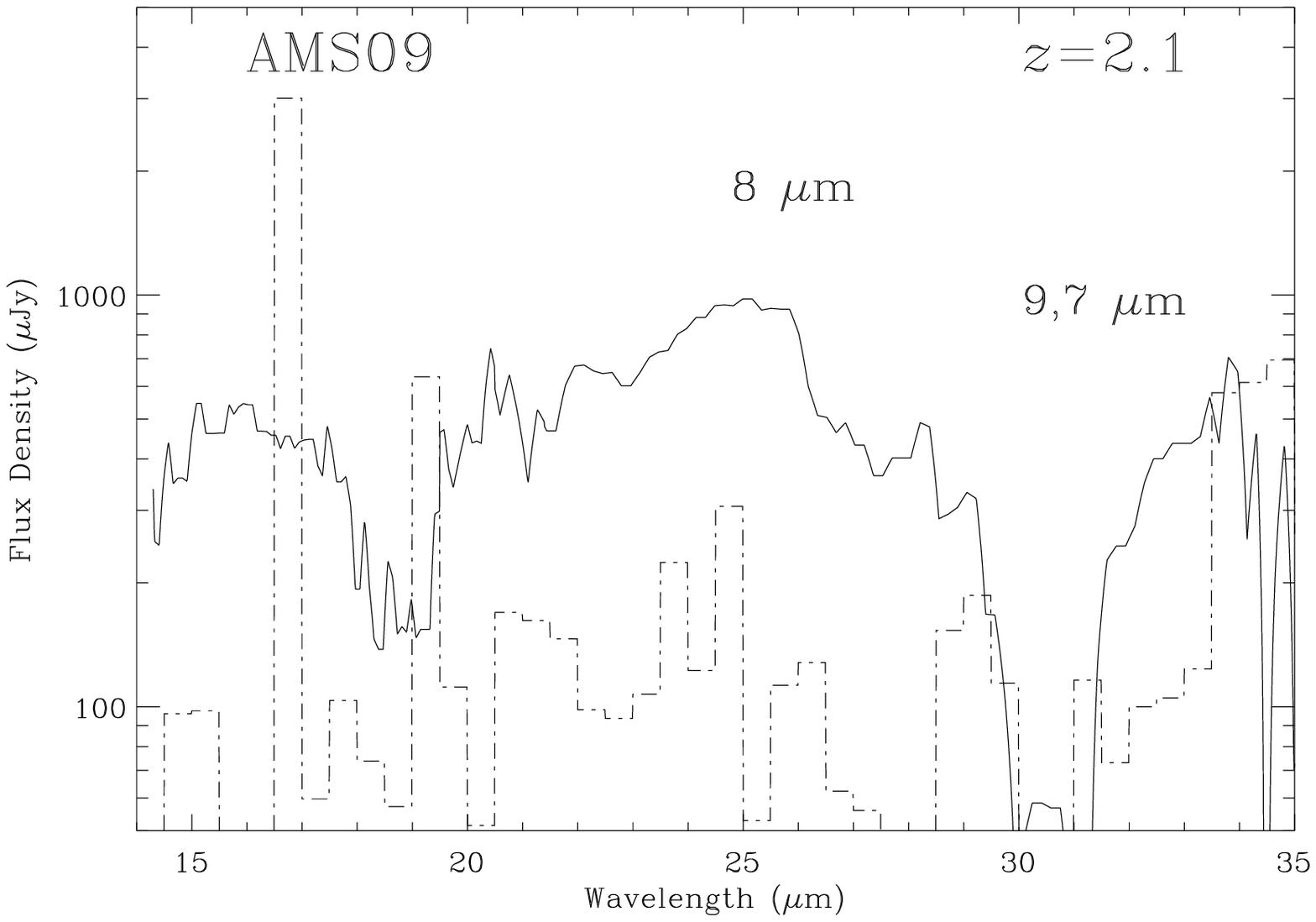} 

\includegraphics{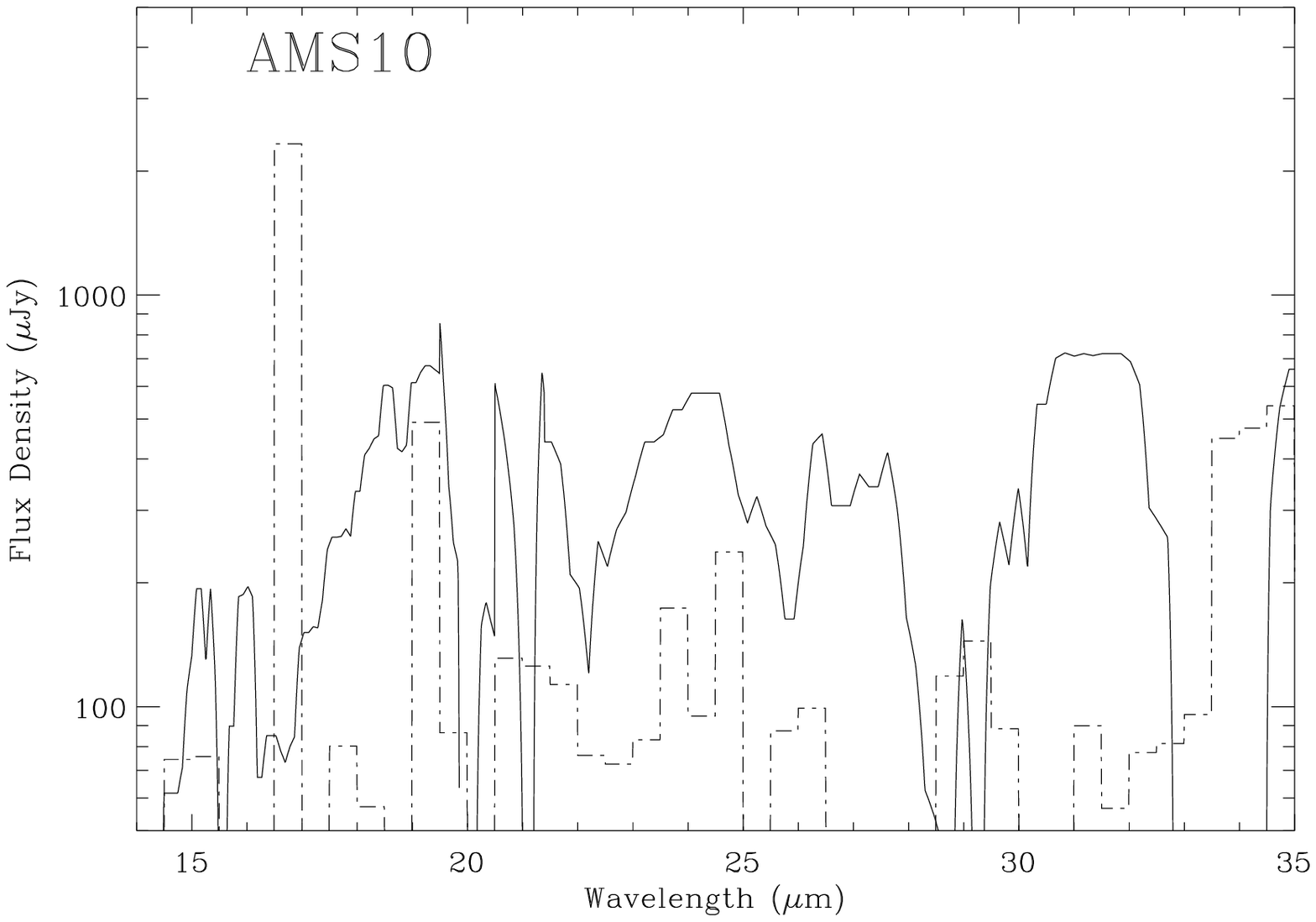} 

\includegraphics{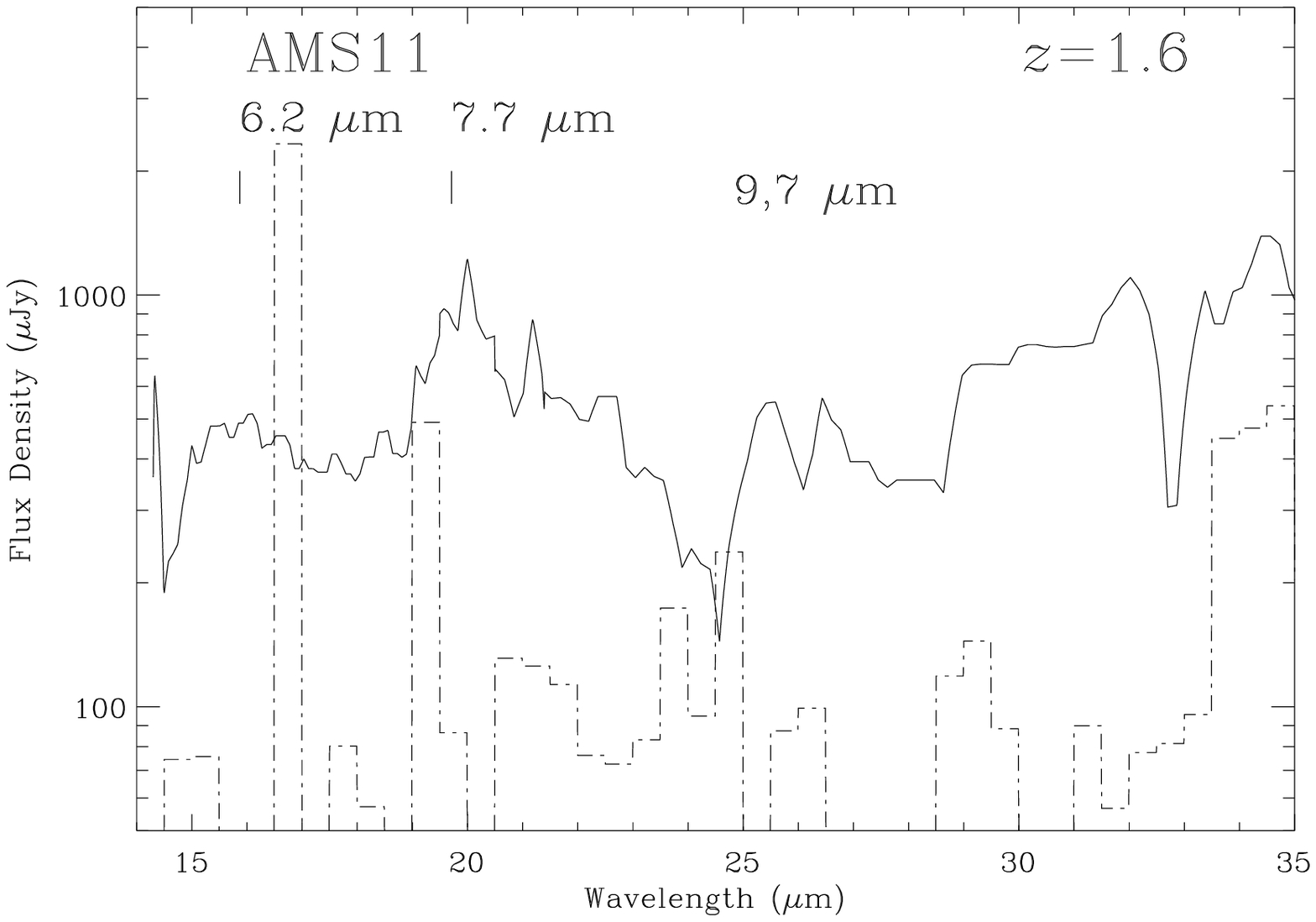} 

\includegraphics{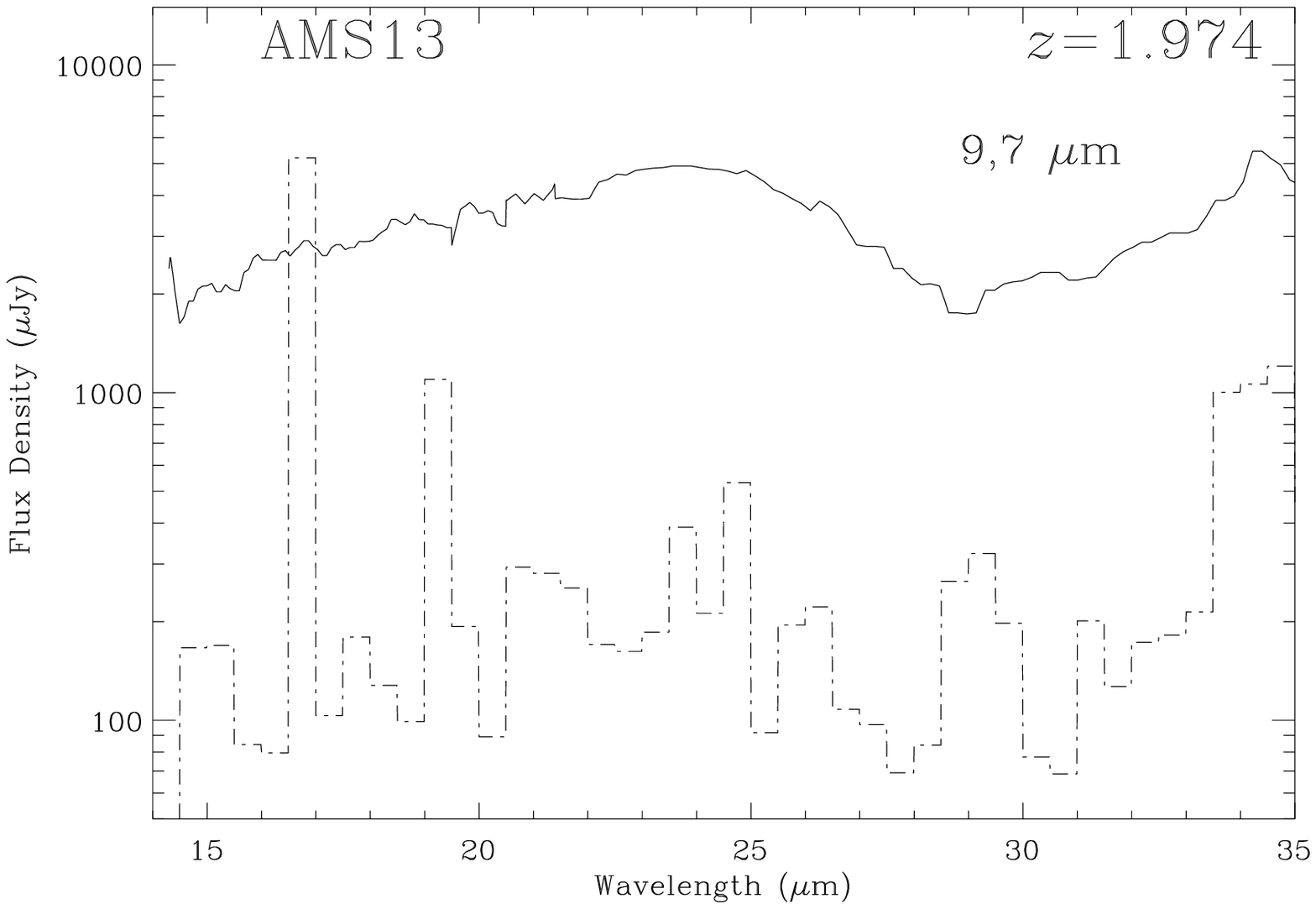} 

\includegraphics{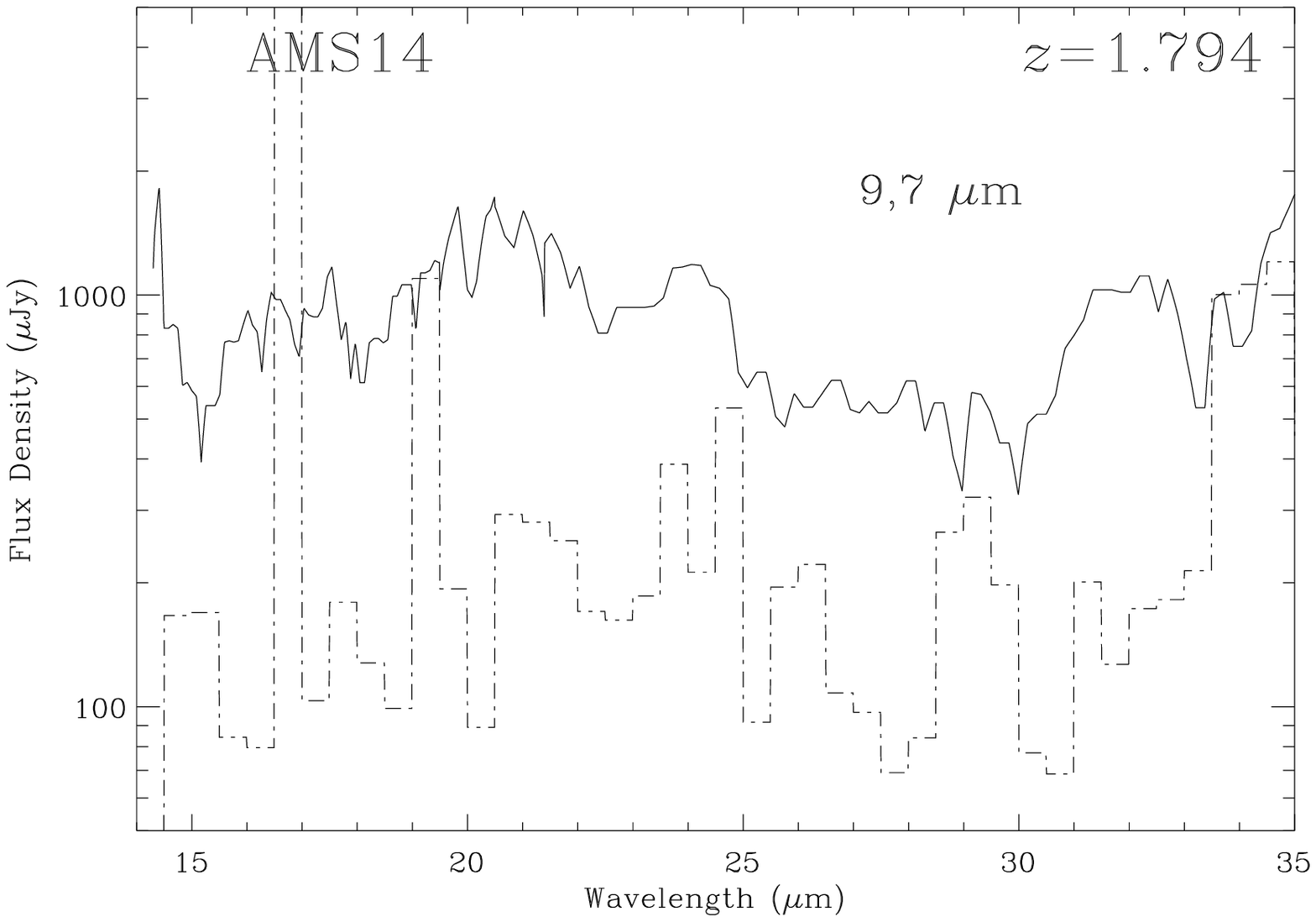} 

{ \caption{ }}
\end{figure*}

\clearpage

\addtocounter{figure}{-1}
\begin{figure*}

\vspace{11cm}

\includegraphics{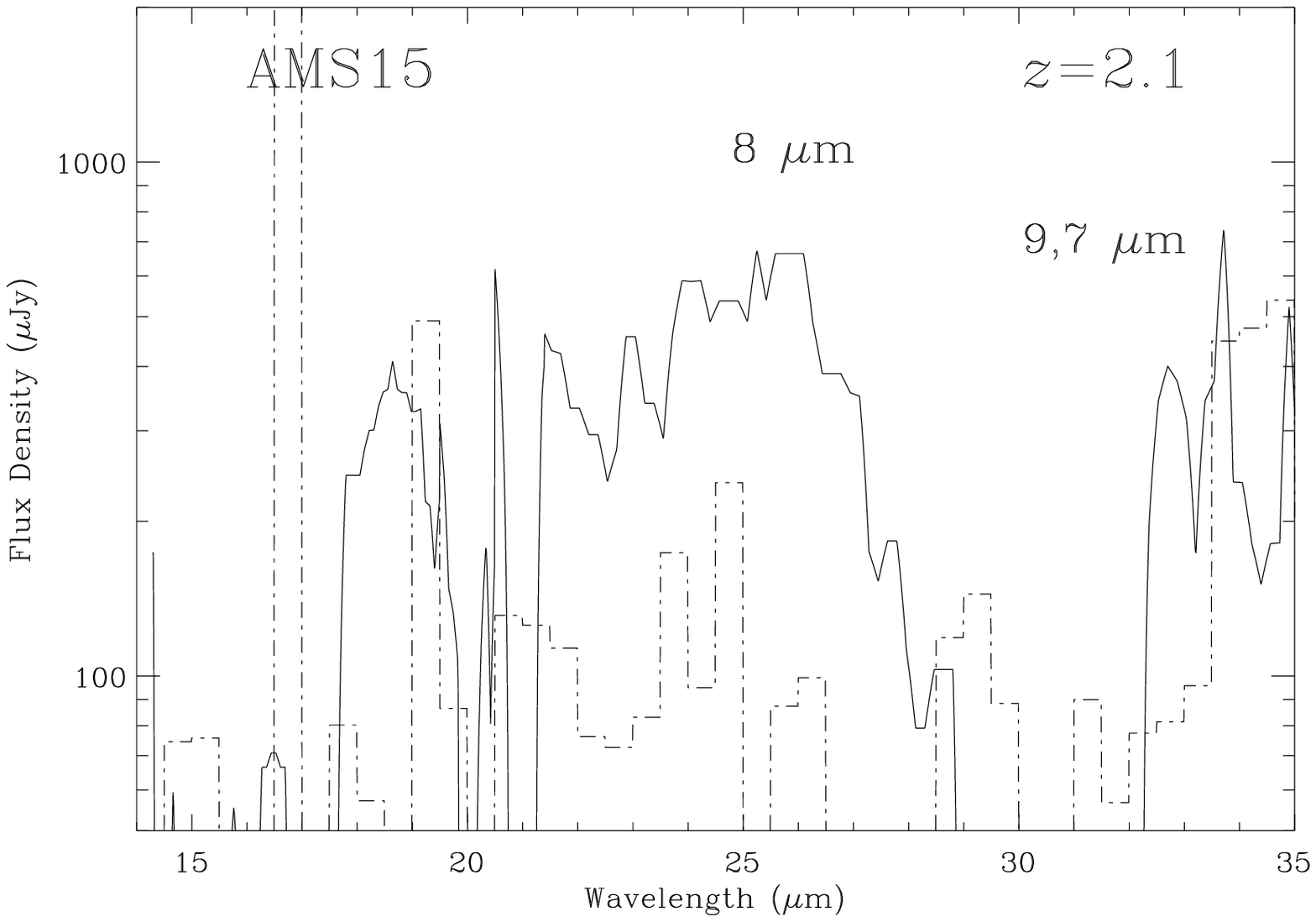} 

\includegraphics{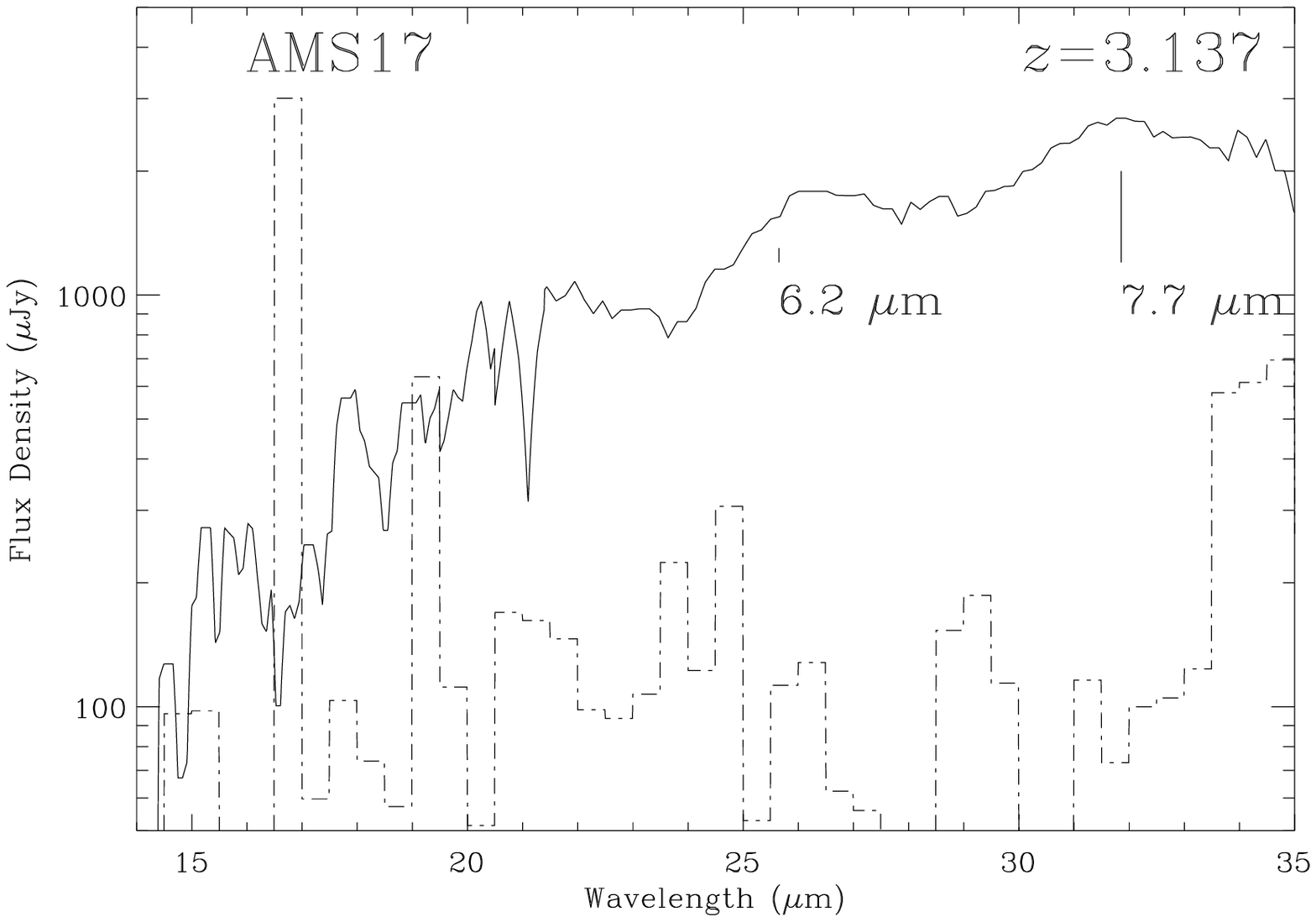} 

\includegraphics{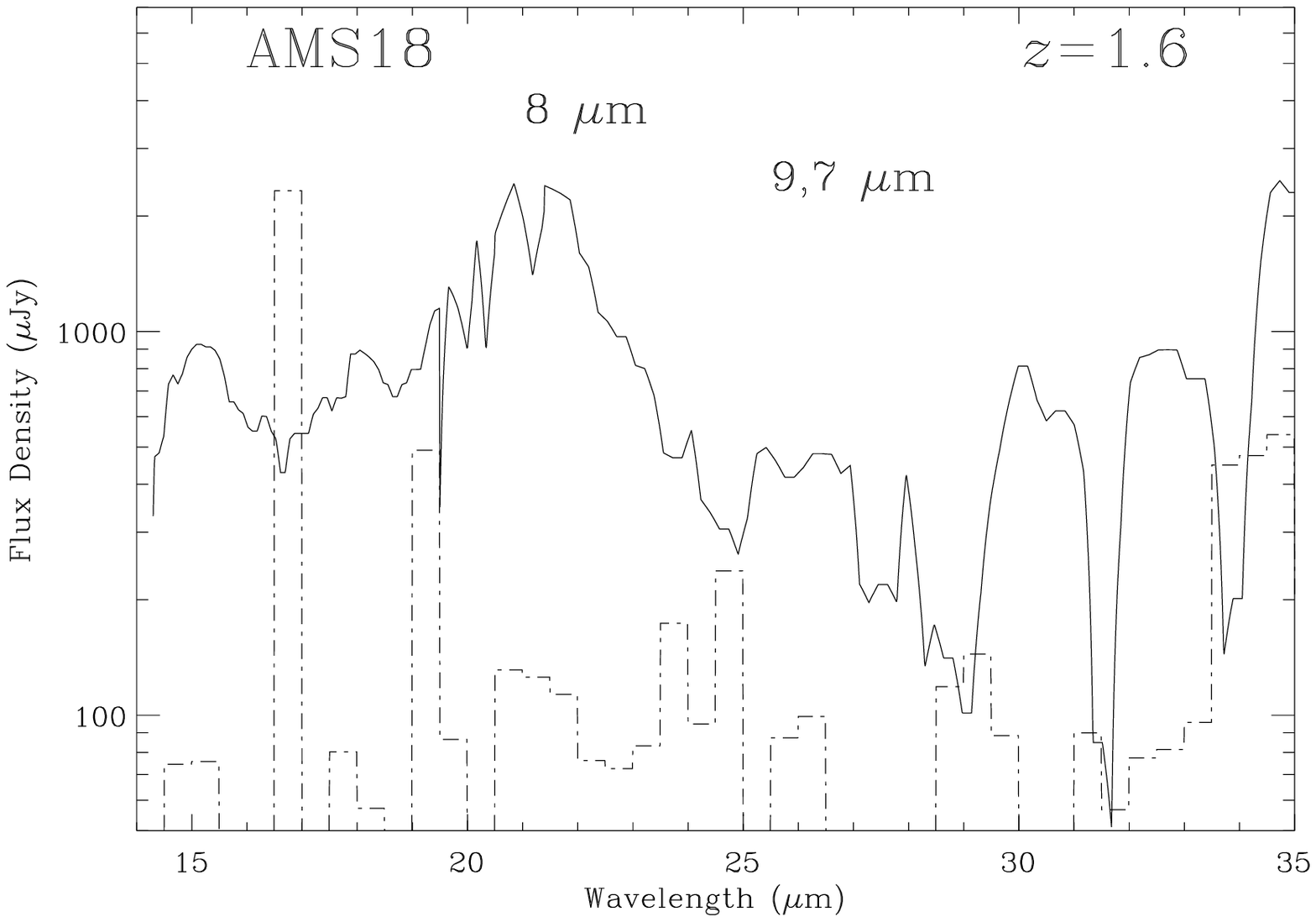} 

\includegraphics{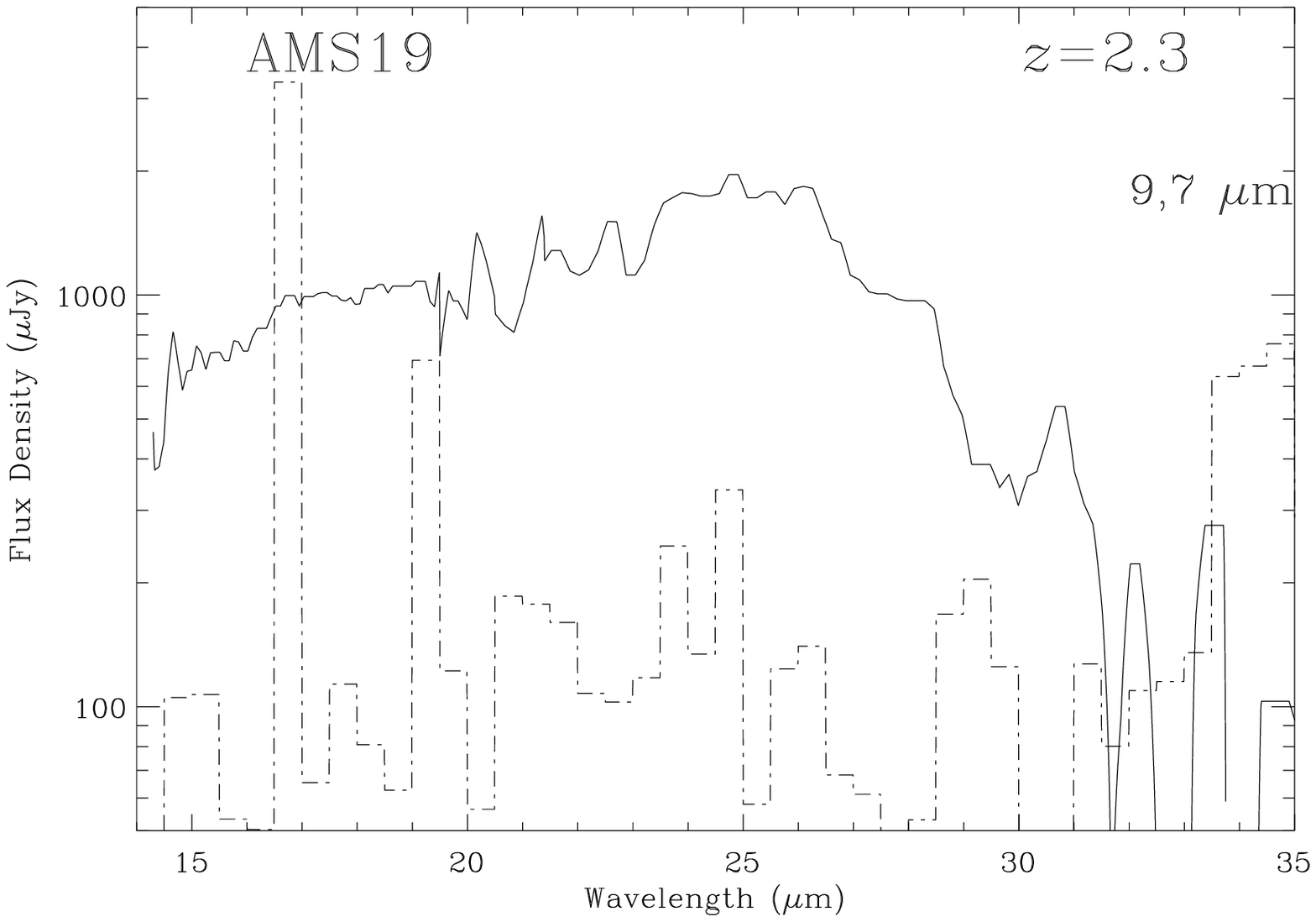} 

\includegraphics{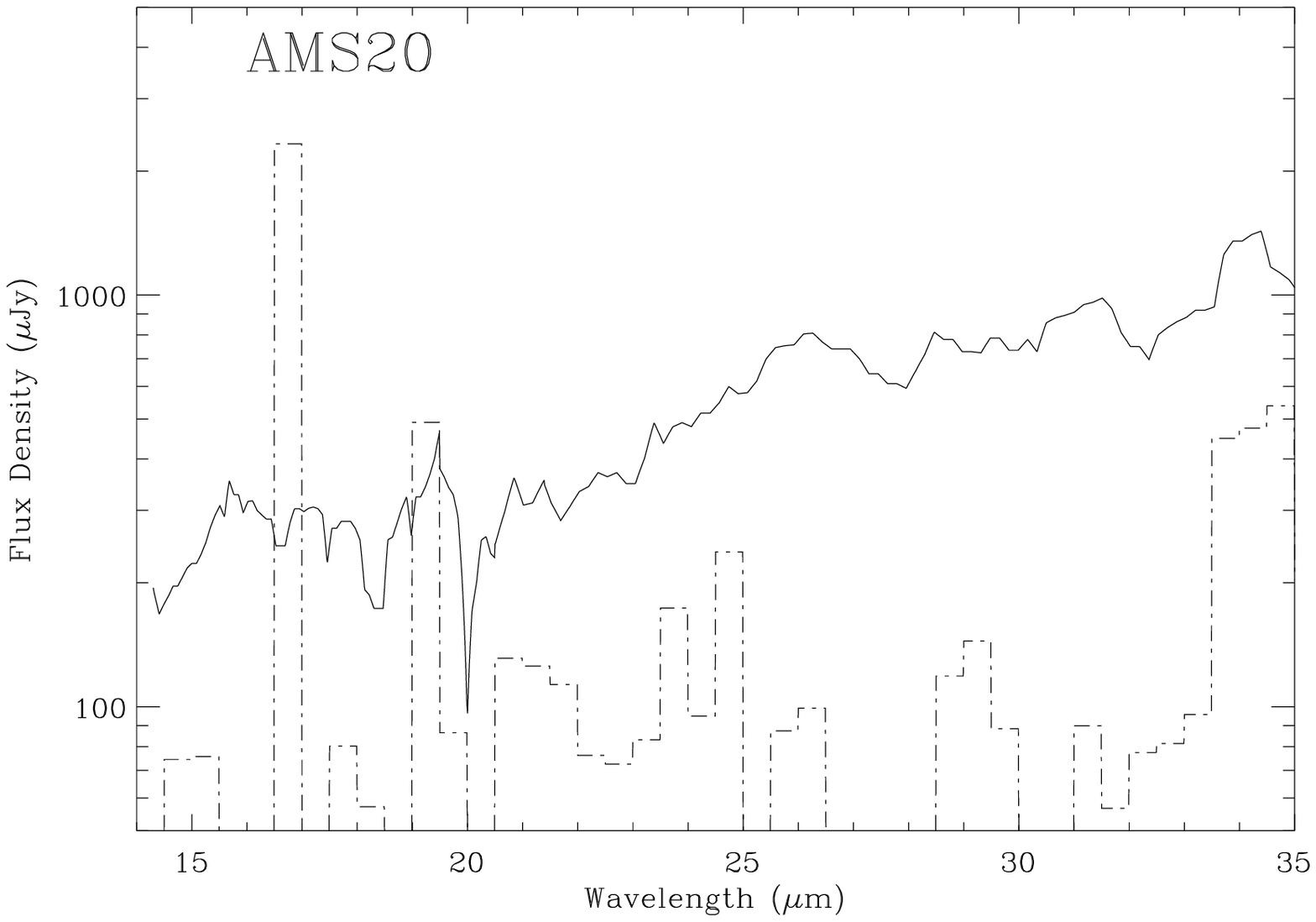} 

\includegraphics{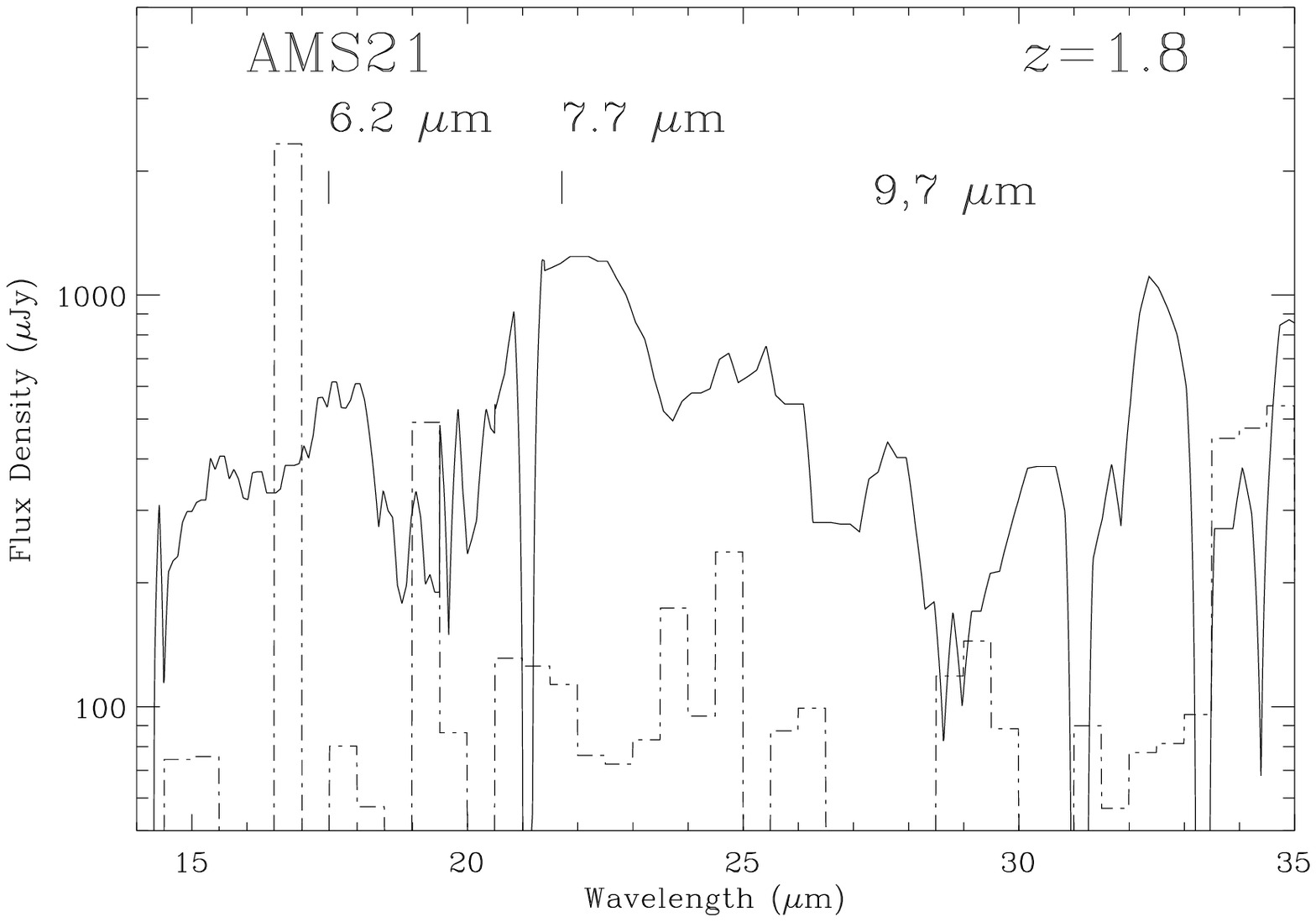} 
\vspace{9cm}

\caption{Individual mid-infrared spectra of all 18 sources observed. As a
  guide to the eye, a the root-mean square noise estimate is plotted as a
dashed-dotted histogram. It was obtained by extracting a spectrum from a
blank region in a 2D spectrum, and scaling it to the appropriate number of
cycles. For sources with a spectroscopic redshift, this is quoted in the top
right corner. The features used to derive a redshift are also indicated at the
observed wavelength. }
\end{figure*}

\clearpage

\begin{figure*}
\begin{center}
\setlength{\unitlength}{1mm}
\begin{picture}(120,145)
\put(0,0){\includegraphics{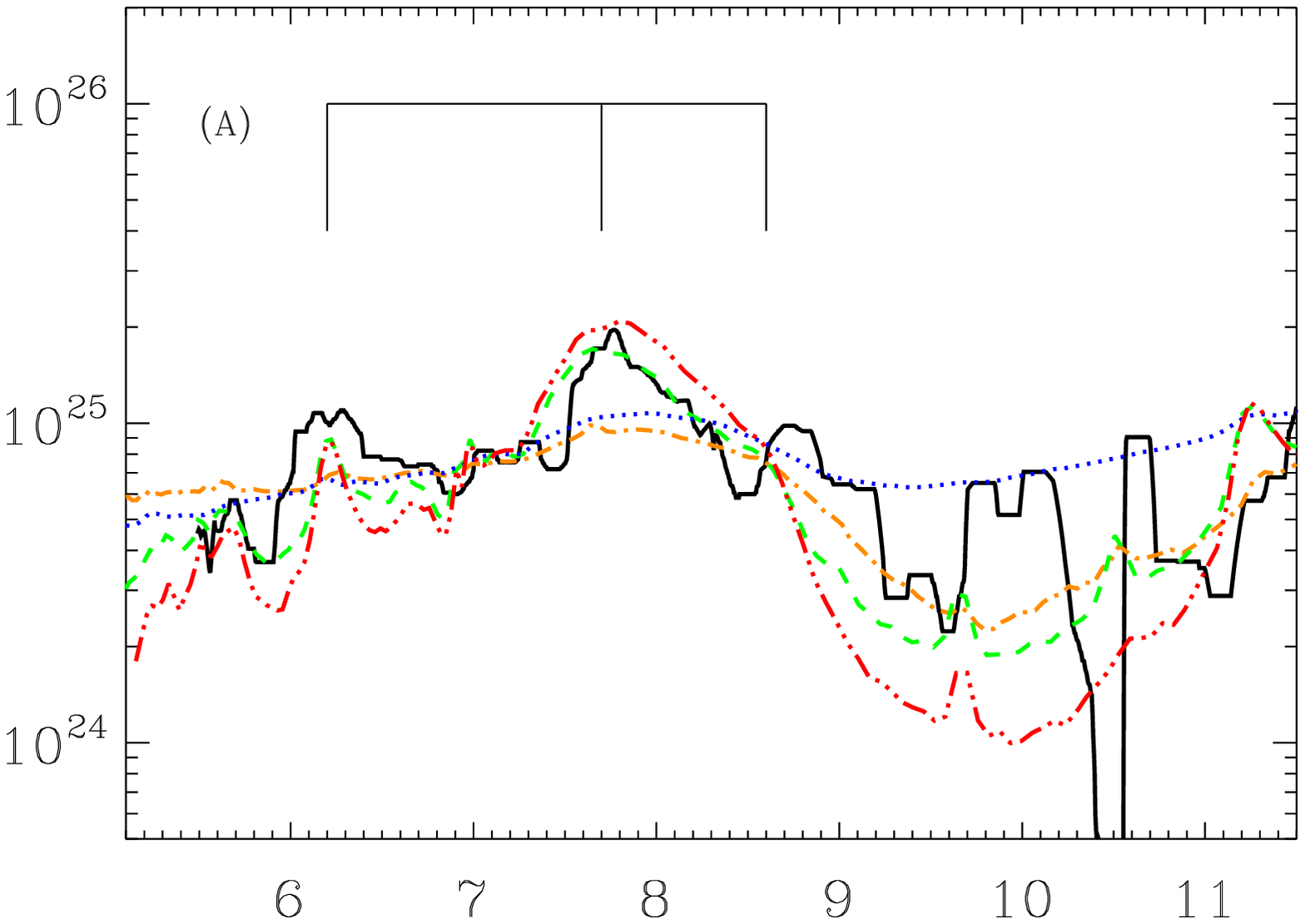}}
\put(0,0){\includegraphics{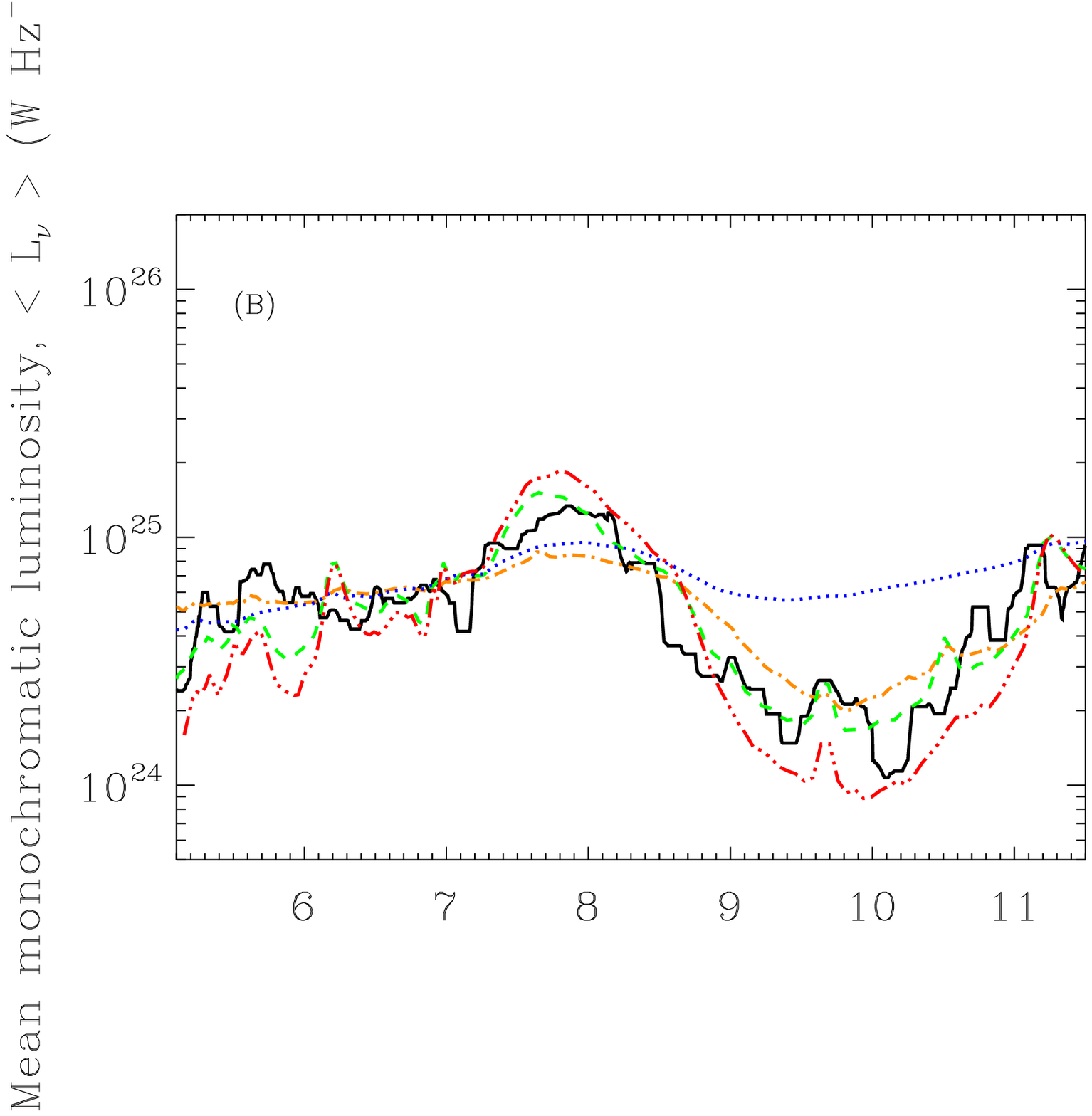}}
\put(0,0){\includegraphics{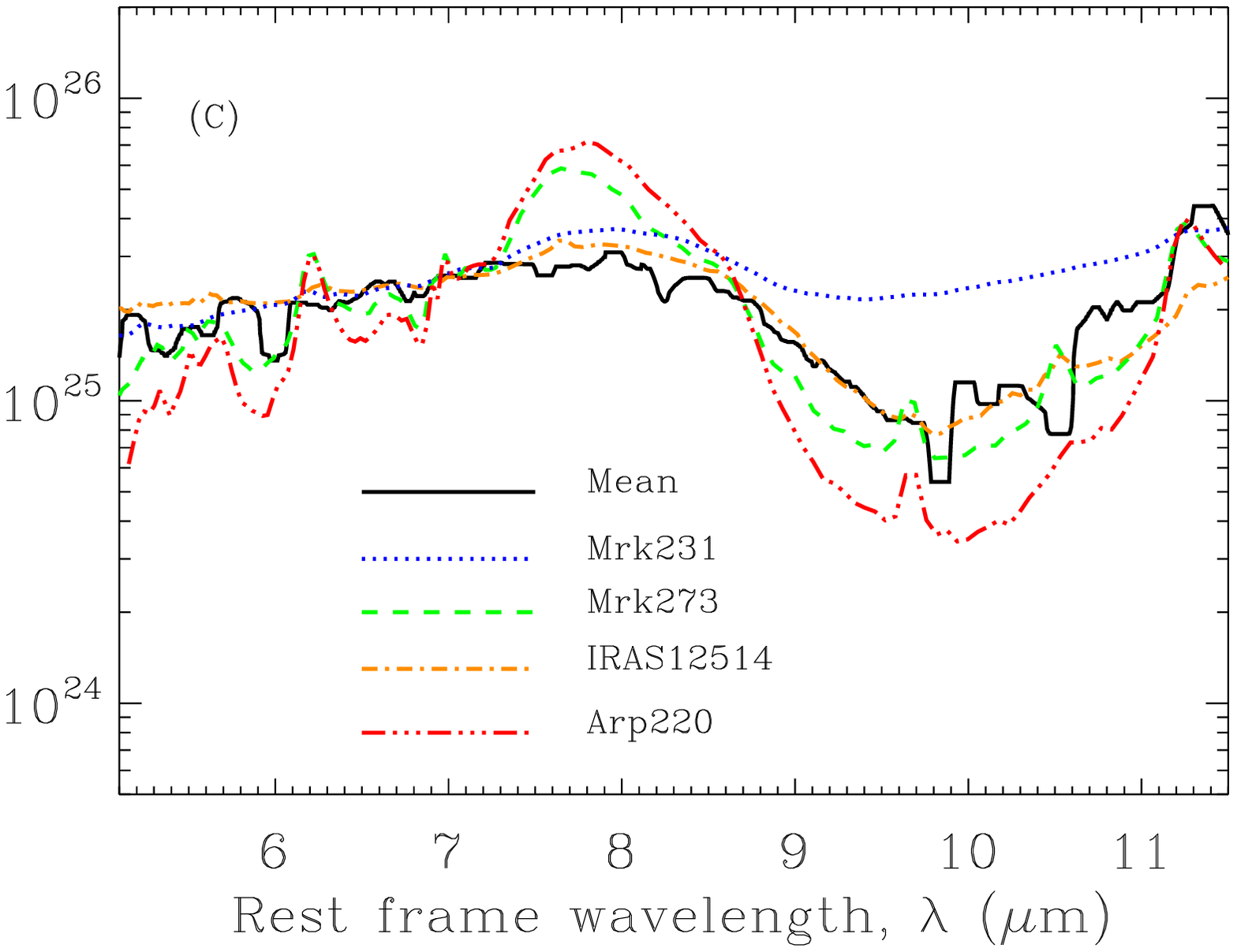}}
\end{picture}
\end{center}
\end{figure*}

\begin{figure*}
\vspace{2cm} { {\caption{ Mean mid-infrared spectra of the
      subsamples.  Overlayed for reference are high-quality low-resolution
      mid-infrared spectra of four local ultra-luminous infrared galaxies:
      Arp~220 (a starburst-dominated ULIRG), Mrk~273 and Mrk~231 (both
      AGN-starburst composites) and IRAS~12514+1027 (with an AGN dominated
      spectrum). These are normalised to go through the data around 7~$\mu$m.
      Subsample~A (top): mean spectrum of 3 sources with both the 7.7 and
      6.2~$\mu$m PAHs visible. The wavelength of these two PAHs, together with
      the 8.6~$\mu$m one, are marked with lines.  Subsample~B (middle): this
      includes 7 objects with an excess around 8 ~$\mu$m, due to the lack of
      6.2~$\mu$m PAH, these are not included in subsample~A. The mean spectrum
      shows the excess emission to be very broad, and the 6.2~$\mu$m PAH is
      still not detected, suggesting this is due to the minimum in optical
      depth around 8~$\mu$m and not the 7.7~$\mu$m PAH. Subsample~C (bottom): this consists of 4 sources
      with clear continuum, and no hint of an excess around 7.7 or 8~$\mu$m.
      The mean spectrum is well described by that of IRAS~12514.} }}

\end{figure*}

\clearpage

\clearpage
\begin{figure}
\begin{center}
\setlength{\unitlength}{1mm}
\begin{picture}(120,130)
\put(0,0){\includegraphics{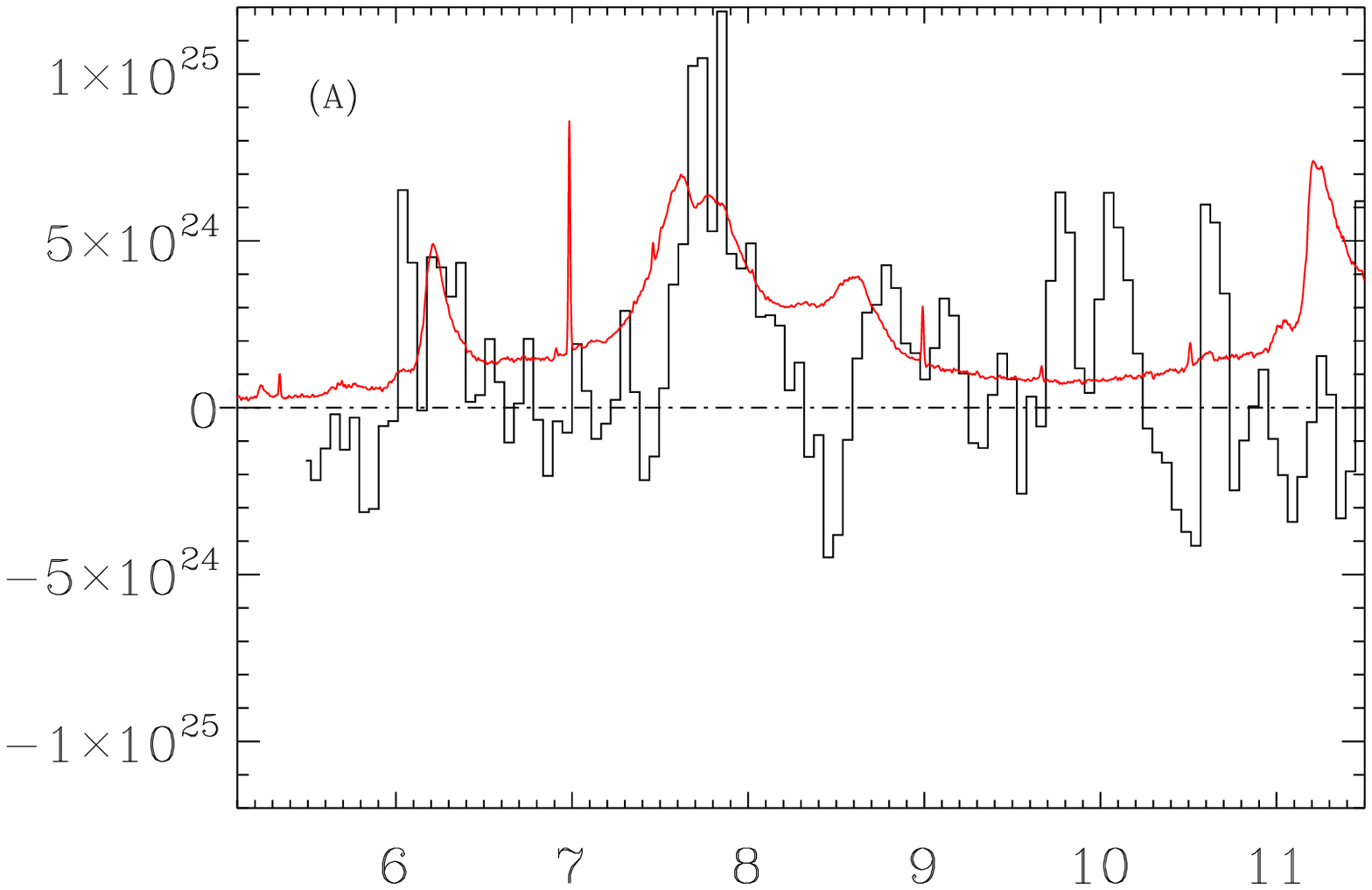}}
\put(0,0){\includegraphics{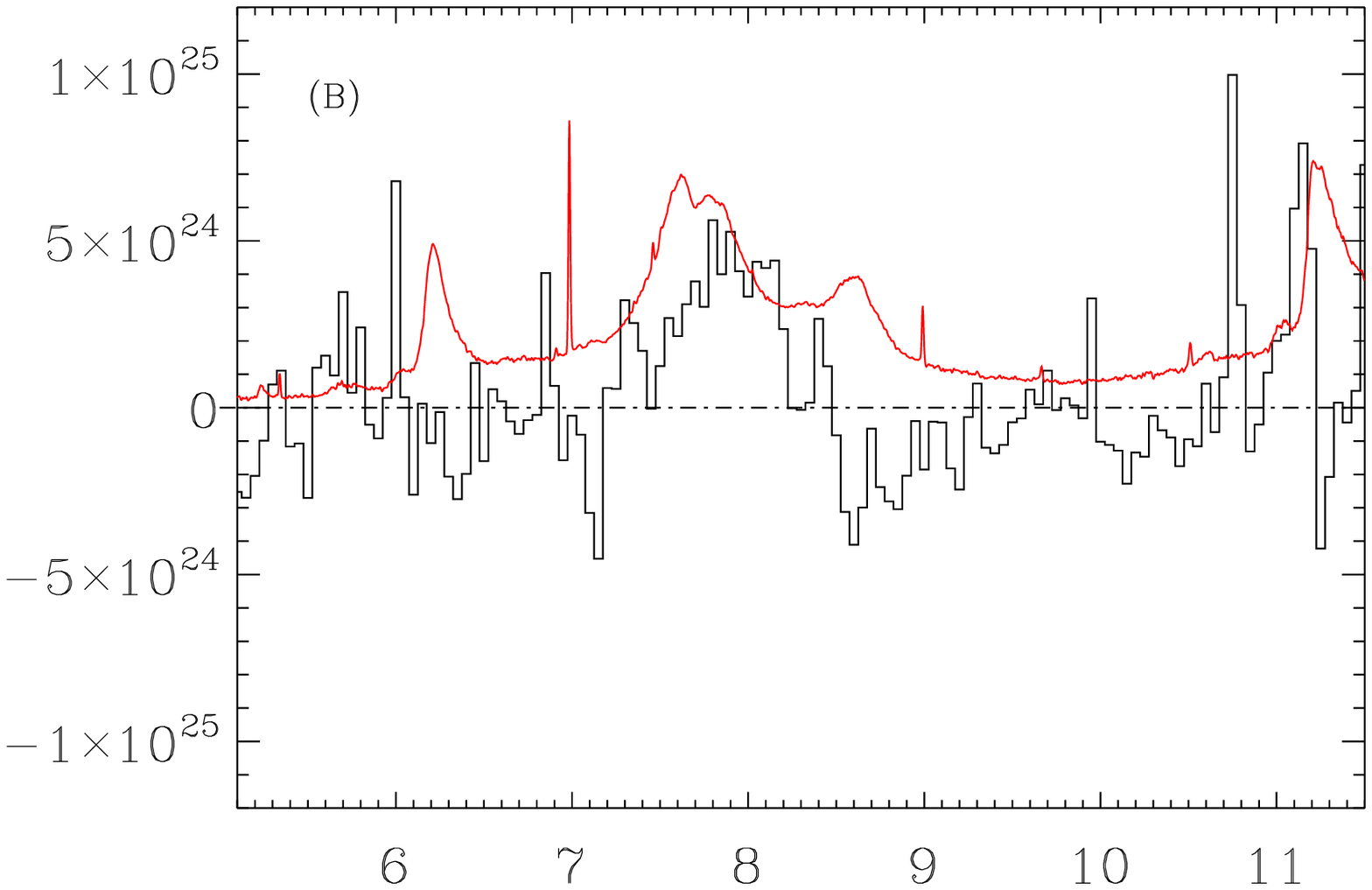}}
\put(0,0){\includegraphics{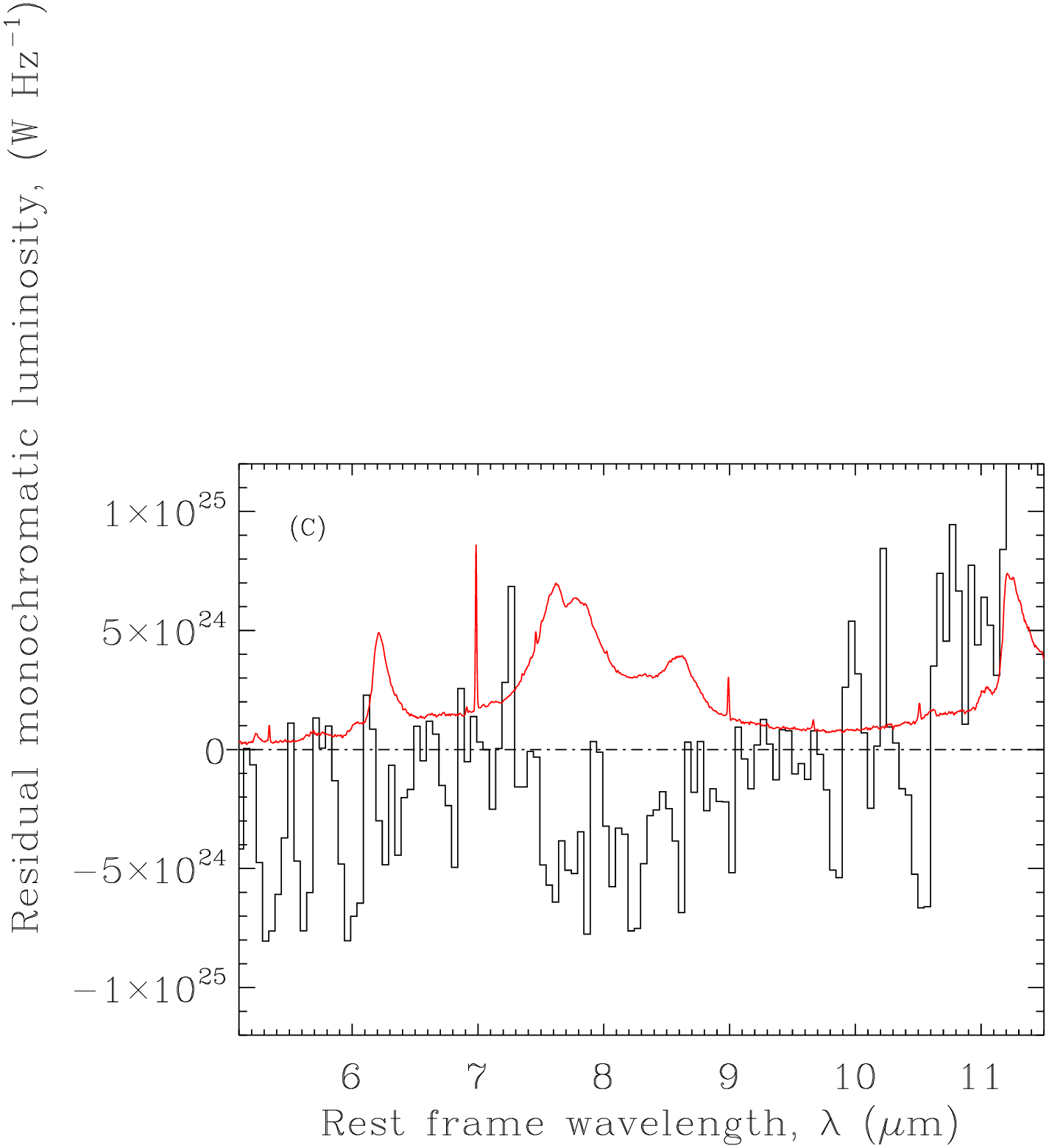}}

\end{picture}
\end{center}
\vspace{0cm} 
 {\caption{ Residuals after subtracting the template of
    IRAS~12514 to the mean spectra of subsamples~A (top), B (middle) and C
    (bottom). For reference, the mid-infrared spectrum of M82 is overlayed,
    which shows the expected positions of the 6.2, 7.7 and 8.6~$\mu$m PAHs.
    Subsample~A has two regions of excess flux, compared to either template,
    these are at the exact wavelengths of the 6.2 and 7.7~$\mu$m PAHs, and
    have the correct width.  Subsample~B shows a broad excess centred at 8
   ~$\mu$m, and no hint of the 6.2~$\mu$m PAH. Although the latter could be
    lost in the noise, the excess at 8~$\mu$m is probably best explained by
    continuum. Therefore, we do not consider subsample~B to have a detection
    of the 7.7~$\mu$m PAH. Subsample~C shows no residual, so the template
    provides a very good description.  The region of 5.5-8.5~$\mu$m has lower
    noise than the region 8.5-11.5~$\mu$m, and of the two regions only the
    former is used to estimate the significance of the detections in
    Table~\ref{tab:pah}. }}
\end{figure}

\clearpage

\begin{figure}
\begin{center}
\setlength{\unitlength}{1mm}
\begin{picture}(150,120)
\put(0,0){\includegraphics{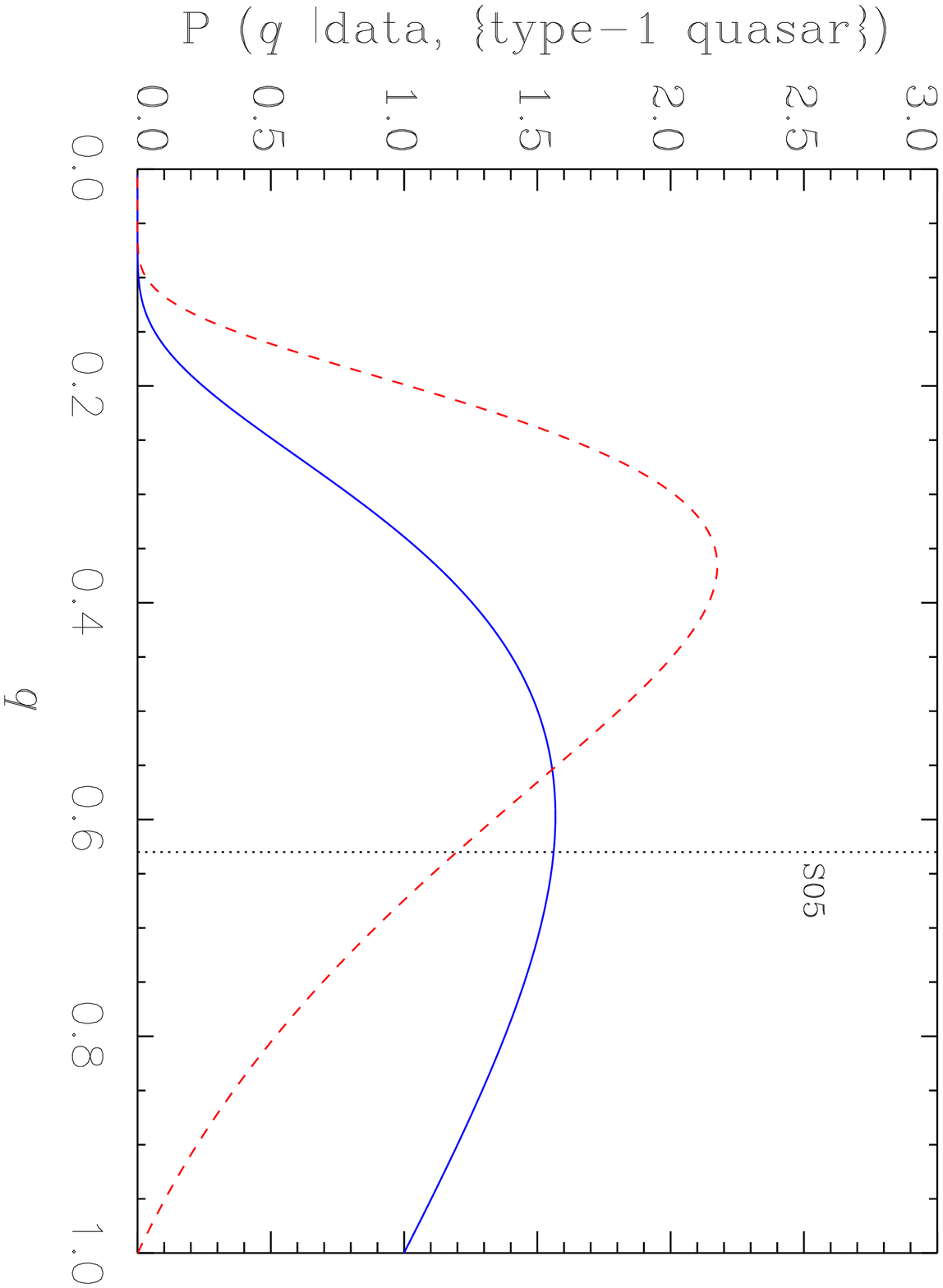}}
\end{picture}
\vspace{-5cm}
 {\caption{Posterior probability distribution function of
    the quasar fraction, $q$, in the redshift range $1.7 \leq z\leq 2.6$,
    given our data and given our modelled number of type-1 quasars. The solid
    blue line shows the distribution if we only use type-2 quasars showing
    lines in the optical spectra (corresponding to rest-frame ultraviolet),
    while the dashed red line shows the case where spectroscopic redshifts
    from Spitzer IRS have been used to complement the optical redshifts. Note
    that only spectroscopically-confirmed type-2 quasars are used. The small
    number of sources used to derive the solid blue curve leads to a large
    error, so that the probability of $q\geq1$ is not zero. For the dashed red
    line, the number of sources used is large enough to avoid this problem.
    The vertical line labelled 'S05' marks the prediction of the Simpson
    (2005) receding torus model, which should only be compared to the solid
    blue line.  } }
\end{center}
\end{figure}

\clearpage

\begin{figure}
\begin{center}
\setlength{\unitlength}{1mm}
\begin{picture}(150,120)
\put(0,0){\includegraphics{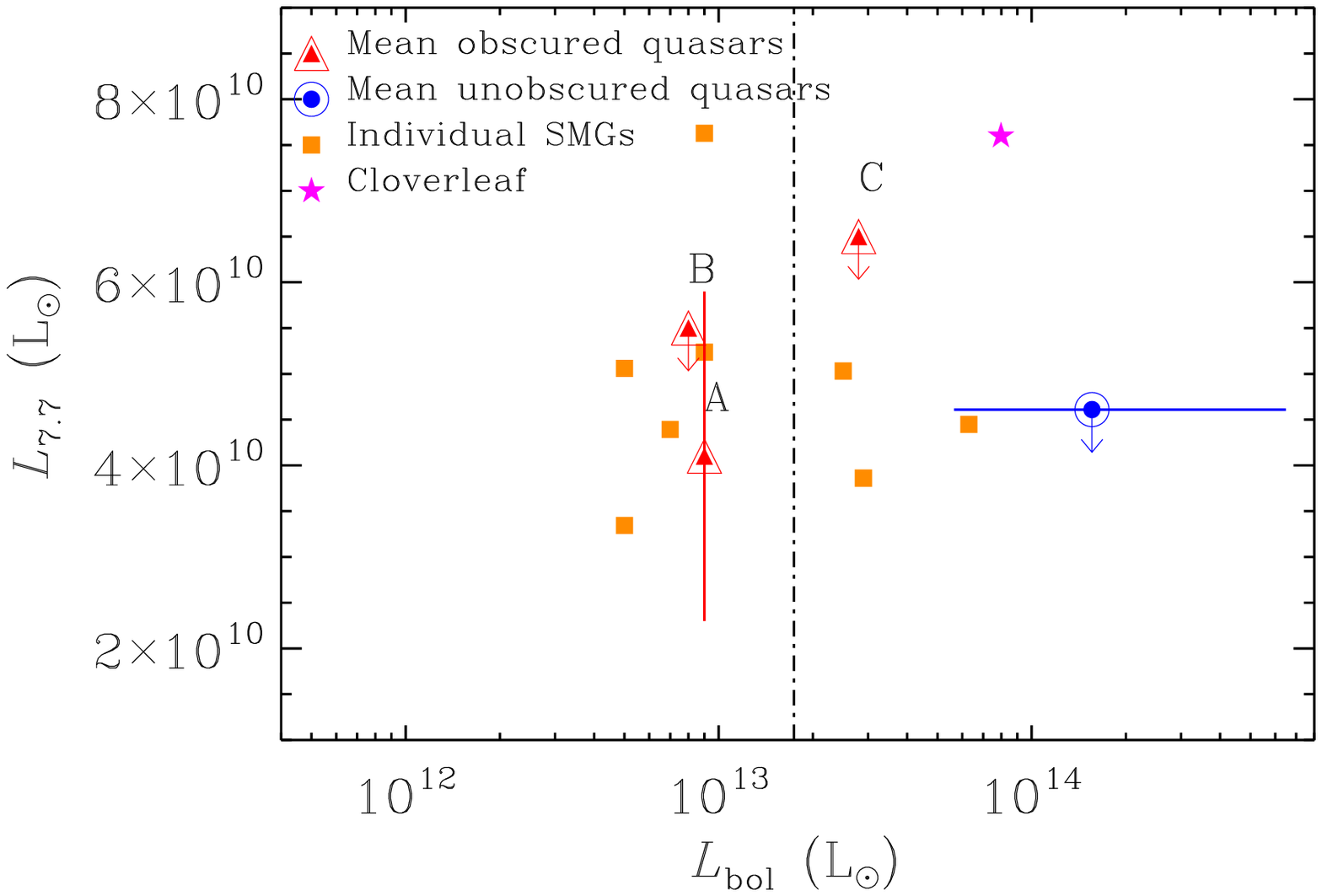}}
\end{picture}
\vspace{-5cm}
 {\caption{Luminosity of the 7.7~$\mu$m PAH line $L_{7.7}$ vs
    bolometric luminosity $L_{\rm bol}$ (both in solar luminosities) for
    samples of $z\sim2$ quasars and submillimetre-selected galaxies (SMGs)
    with comparable mid-infrared spectroscopy.  Upper limits are 3~$\sigma$.
    Our sample of $z\sim2$ obscured quasars is represented by two points: the
    mean value for detections (Figure~2 panel~A) and the upper limit from the
    non-detections (Figure~2 panels~B and C).
    The unobscured quasars are from the sample of \citet
    {2007A&A...468..979M}, the ``Cloverleaf''quasar is from
    \citet{2007ApJ...661L..25L}, the SMGs from \citet{2007ApJ...660.1060V}.
    The vertical dashed line represents the break of the
    unobscured quasar luminosity function \citep[LF][]{2004MNRAS.349.1397C},
    converted to $L_{\rm bol}$ using the SED of \citet{1994ApJS...95....1E}.
    This should only be compared to quasars (both obscured and unobscured) but
    not to SMGs. } }
\end{center}
\end{figure}

\end{document}